\documentclass{aa}
\usepackage{graphics}
\begin{document}

%   \thesaurus{11     % A&A Section 11: galaxies
%              (11.01.2;
%               11.02.1;
%               11.07.1;
%               11.10.1;
%               11.17.3;
%               11.19.1)}
%%%%%%%%%%%%%%%%%%%%%%%%%%%%%%%%%%%%%%%%%%%%
   \title{A new list of extra-galactic radio jets}

   \author{F.K. Liu\inst{1,2} 
	\and
	   Y.H. Zhang\inst{2}}

   \offprints{F.K. Liu}

   \institute{Department of Astronomy and Astrophysics, G\"oteborg
	University \&
	Chalmers University of Technology, 41296
	G\"oteborg, Sweden \\
	email: fkliu@fy.chalmers.se
	\and 
	International School for Advanced Studies,
	Via Beirut 2-4, 34014 Trieste, Italy
             }

   \date{Received 1 June, 2000; accepted 1 June, 2000}

   \abstract{
	A catalogue of extra-galactic jets is very useful both in 
	observational and theoretical studies of active galaxies.
	With the use of new powerful radio instruments, the detailed 
	structures of very compact or weak radio sources are investigated
	observationally and many new radio jets are detected. In this
	paper, we give a list of 661 radio sources with detected radio
	jets known to us prior to the end of December 2000. All references are
	collected for the observations of jets in radio, IR, optical,
	UV and X-ray wave-bands. 

	}
   \maketitle

%
%________________________________________________________________

\section{Introduction}

Theoretically every extra-galactic radio source contains at its center
an accretion disc surrounding a super-massive black hole and ejecting
two symmetric relativistic motion plasma jets in opposite directions 
along its rotation axes (Begelman et al. \cite{Begelman}). Whether a 
jet can be detected depends not only on the properties of the radio 
source itself but also on the instruments used. Radio jets have been 
detected in various types of radio sources: FR I and FR II radio 
galaxies, Seyfert galaxies, radio quasars, and BL Lac objects (Bridle 
\& Perley \cite{Bridle3}; Liu \& Xie \cite{Liu} hereafter Paper I), 
though the detected rates of jets are different in 
different categories of radio sources with the same instruments
under similar working conditions. The different detected rates of jets
can be reconciled with the unification model of active galaxy nuclei
(AGNs) (Bridle \& Perley
\cite{Bridle3}; Antonucci \cite{Antonucci}). By studying jets, one can
learn the physics of a jet itself, the ambient medium, and the central
engine of the radio source. Many observational works have been
dedicated to the studies of the structures of galaxies 
(e.g. Leahy et al. \cite{Leahy} and references therein; Owen \& Ledlow
\cite{Owen} and references therein; Xu et al. \cite{Xu} and references
therein, Morganti et al. \cite{Morganti}; Condon et al. \cite{Condon};
Parma et al. \cite{Parma}; Carilli et al. \cite{Carilli}), radio 
quasars and BL Lac objects (Lonsdale et al. \cite{Lonsdale}; Murphy 
et al. \cite{Murphy}; Bridle et al. \cite{Bridle2}; Price et al. 
\cite{Price}; Akujor et al. \cite{Akujor}; Garrington et al. 
\cite{Garrington}; Kollgaard et al. \cite{Kollgaard};
Laurent-Muehleisen et al. \cite{Laurent}) since the publication of the
jet list given by Liu \& Xie (\cite{Liu}). These works describe a
large number of kpc-scale jets in many radio sources. With Very Large
Baseline Array (VLBA) and global  
Very Long Baseline Interferometer (VLBI), the structures of the
center of many radio sources are investigated in detail and pc-scale
jets (e.g. Spencer et al. \cite{Spencer}; Taylor et al. \cite{Taylor}
and references therein; Dallacasa et al. \cite{Dallacasa}; Fey \&
Charlot \cite{Fey} and references therein) are revealed in many 
compact steep spectrum radio sources and flat-spectrum radio
sources. A recent review is given by Zensus (\cite{Zensus}) on the 
understanding of the properties of pc-scale radio jets in
extra-galactic radio sources. 

In their paper, Bridle and Perley (\cite{Bridle3}) list 125 confirmed
and 73 possible 
extra-galactic radio jets. In 1992, Paper I gave 276 radio sources with
detected jets known to them prior to mid-December 1989. As many more 
detected extra-galactic radio jets have been reported in the 
literature since then, it is time to give a updated list of radio jets. 
To make the new list in this paper consistent with those reported in the
literature, we adopted the morphological definition of a
radio jet given by Bridle and Perley (\cite{Bridle3}). In addition to the
information on luminosities of jets, total sources and central cores,
the length, sidedness and simple code of morphology were also given in
this work. A detailed description of 
the catalogue will be given in Sect.~2. 
A simple discussion and conclusion will
be given in Sect.~3. Jet properties in IR, 
optical, UV, and X-ray bands will be discussed later (Liu et al. 
\cite{Liub}).  A Friedmann standard cosmology model is adopted and
Hubble constant $H_0 = 100 Km \cdot s^{-1} Mpc^{-1}$ and deceleration 
factor $q_0 = 0.5$ are used throughout this paper. 

%__________________________________________________________________

\section{Description of the catalogue of known radio jets}

To distinguish a jet from other narrow features or bridges, we used, as
in Paper I, the criteria given by Bridle and Perley (\cite{Bridle3})
and Bridle (\cite{Bridle1}) as has been adopted often in the
literature: a ``jet'' is a narrow feature that must be

(a) at least four times as long as it is wide,

(b) separable at high resolution from other extended structures (if
any) either by brightness contrast or spatially, and

(c) aligned with the radio nucleus of the parent object where it is
closest to it.

Under this definition of a jet, some sources reported to have detected
jets, e.g. some core-jet sources observed with VLBI, are not included 
in our catalogue. 

The term ``core'' used in Bridle and Perley (\cite{Bridle3}) and paper I 
is resolution-dependent. Bridle et al. (\cite{Bridle2}) define a
``resolution-independent'' term ``central feature'' as an unresolved 
feature coinciding to within observational errors with the best available
position for optical counterparts. It is not always true that a
resolution high enough to resolve an opaque feature can be reached in
the observations and therefore a ``central feature'' is still 
resolution-dependent. Therefore, we still used the term ``core'' as in paper
I. The usual diagnostics for identification of the nucleus is the
positional coincidence with the optical object and flat or inverse
radio spectra. Additional constraints are often found from evidence of
variability and polarization characteristics. To resolve as clearly
as possible the extended emission from the core, we use observational data
at a high frequency. 

Table 1 lists 661 radio sources with detected radio jets known to
us prior to the end of December 2000. All jets in the table were
collected from the literature available to us, based on the above 
definitions. In Table 1, column 1 gives the IAU name of the radio source
in order of right ascension at 1950.0 and 
the jet position angle at the nearest to the core -- the letters N, E,
S, W, NE, SE, NW, and SW represent north, east, south, west, north-east,
south-east, north-west, and south-west, respectively. We used the
position angle to identify the jets in a radio source. If the
difference of position angles of  a jet on pc- and kpc- scales
is significant, we used the one detected on the
kpc-scale. If the jet is one-sided (for the definition of sidedness, see
the explanation of column 8) or one jet is significantly brighter than
the other one in two-sided sources, the letters C-J follow the counter 
(weaker) jet. If two or more radio galaxies with detected jets are
found in a cluster with one IAU designation, we use letters A and B to
label them according to their right ascension. Many radio sources are
well-known in the literature by their alias name. Column 2 lists
some of those names. 

Column 3 gives the optical identification of a source -- Seyfert galaxies (S),
radio galaxies (G) other than Seyferts, radio quasars (Q), or BL Lac 
object (BL) -- and its red-shift. A red-shift given in brackets is
uncertain. Fifty-three radio sources have no identification in the
table. We identify Seyfert galaxies,
based on the catalogue of quasars and active nuclei
given by V\'eron-Cetty and V\'eron (\cite{Veron}). The numbers 1, 2, and 3
imply type 1, 2, and 3 Seyferts. Identification and red-shift can
usually be found in the literature reference given in the column 10
or the references therein. However, some sources have no identification and 
measured red-shift when the radio observations are done. We find the
information of the identification and red-shift from the following papers or 
catalogues: Hewitt and Burbidge (\cite{Hewitt}), Junkkarinnen et al. 
(\cite{Junkkarinnen}), V\'eron-Cetty and V\'eron (\cite{Veron}), de
Vaucouleurs et al. (\cite{Vaucouleurs}), Sandage and Tammann 
(\cite{Sandage}). The red-shift of 1826+796 is taken from Henstock et 
al. (\cite{Henstock}). The radio source 3C22 was classified as a radio
galaxy in the original survey and is re-classified as a quasar
by Rawlings et al. (\cite{Rawlings}). 

Columns 4, 5 and 6 list, in the observer's frame, the jet, total and core
powers -- $\lg P_j^{1.4}$ at 1.4 GHz, $\lg P_t^{1.4}$ at 1.4 GHz and
$\lg P_c^5$ at 5 GHz, respectively. We used the values at these
frequencies because 
the flux densities $S_j^{1.4}$, $S_t^{1.4}$, and $S_c^5$ values are
most often available. Images at 5 GHz have higher resolution than at
1.4 GHz and thus give flux of a core with less contribution from 
extended structures. However, the images at 1.4 GHz give higher
sensitivity and less observational missing of radio flux. 
We try to use integrated flux densities for jets with a correction 
to background (see Bridle et al. 1994 for detail), if possible. When
the total flux density of a source is not given in the observational
papers of jets or the references therein, we took it from the catalogue 
by V\'eron-Cetty and V\'eron (\cite{Veron}). We estimated core flux,
if not available.  We use ``typical'' values for 
variable total and core emission. Some $S_j^{1.4}$, $S_t^{1.4}$, and 
$S_c^5$ values are estimated with extrapolation or interpolation
from neighboring frequencies, assuming power-law spectra with a spectrum
index of $\nu^{-0.65}$ for the jets, $\nu^{-0.7}$ for the total
emission and $\nu^0$ for the cores if no spectrum information is available.

In column 7 is given the projected length of each jet, $d_j$, in
kilo-parsecs, measured from the core (nucleus) over all the jet
region. 

Column 8 classifies the sidedness of the outer 90\% of the jets. We
adopted the same definition as in Bridle (\cite{Bridle1}) and Bridle 
and Perley (\cite{Bridle3}): one side ``1'', two side ``2'', and transition 
``T'', according to the ratio of luminosities between the brighter and
fainter jets, larger than 4, less than 4, or intermediate. The ratio is 
usually higher near the core and becomes smaller towards the end of a jet.
There is no special physical reason to use 4 instead of other values.

%
%__________________two column table 3________________
\begin{table*}
	\begin{description}
	\item[Table 2.] Statistics of Table 1
	\end{description}
\def\hline{\noalign{\hrule}}
\tabskip=0pt
\offinterlineskip
\halign{
\strut\hskip3pt#&\hskip3pt#&\hfill#\hfill& \hfill#\hfill& \hfill#\hfill&
\hfill#\hfill& \hfill#\hfill& \hfill#\hfill& \hfill#\hfill& \hfill#\hfill&#\cr
\noalign{\hrule height 1pt}
\noalign{\smallskip}
&&\multispan2\hfil Z\hfil&\multispan2\hfil $\lg P_j^{1.4}$\hfil&\multispan2
\hfil $\lg P_t^{1.4}$\hfil &\multispan2  \hfil $\lg P_c^5$\hfil&\cr
\hfill  Type \hfill&\hfill Number\hfill&\multispan2\hrulefill&\multispan2
\hrulefill&\multispan2\hrulefill&\multispan2\hrulefill&\cr
&&\hfil\quad Min\quad \hfil&\hfil \quad Max\quad \hfil &\hfil\quad  Min\quad 
\hfil &\hfil \quad Max\quad \hfil &\hfil\quad Min\quad \hfil &\hfil \quad Max
\quad \hfil&\hfil \quad Min\quad \hfil &\hfil \quad Max\quad \hfil&\cr
\noalign{\hrule}
\noalign{\smallskip}
%                      Z        $P_j^{1.4}$  $P_t^{1.4}$    $P_c^5$
%Source	 Number	 ------------  -----------  -----------  -----------
%		   Min    Max    Min    Max    Min    Max    Min   Max
%==========================================================================
QSOs	&  243&  0.104 & 3.886& 24.74& 27.94& 25.34& 28.76& 23.67& 28.48&\cr
G	&  279&	 0.0012& 3.395& 20.99& 27.84& 21.68& 27.94& 20.38& 27.09&\cr
S	&  50 &	 0.002 & 0.518& 20.54& 25.42& 21.34& 27.92& 19.28& 25.54&\cr
BL	&  36 &	 0.0177& 1.404& 22.15& 27.44& 23.68& 28.15& 23.41& 28.20&\cr
U	&  53 &	 ---   &---   &  --- &---   & ---  &---   &---   &---   &\cr
\noalign{\hrule height 1pt}
}
\end{table*}

Codification of the morphology of a jet is given in column 9 -- C-symmetry
(C), S-symmetry (S), L-shape (L), very complicated (B) and almost
straight (blank). Jet morphology provides information about a jet, 
interaction between a jet and the intergalactic medium (IGM), the central
engine and its motion in the IGM. Narrow-angle tail
sources usually have jets of C shape.

Original references reporting the detection of a jet in radio, 
infrared, optical, ultraviolet, and X-ray wave-bands are given in
column 10. Bridle and Perley (1984) is always given as one of the
references, if the source is listed in their catalogue. 

\begin{figure}
   \resizebox{\hsize}{!}{
	\includegraphics{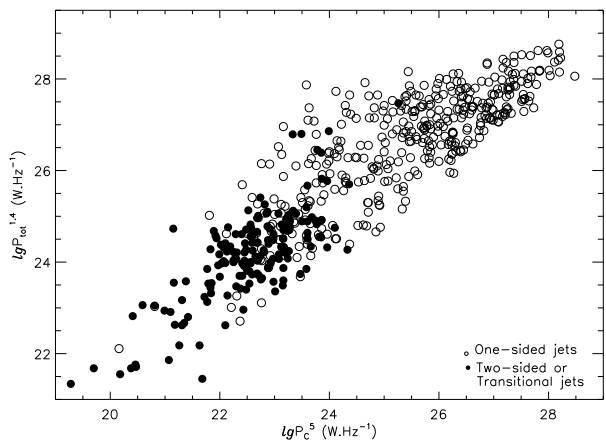}
	}
	\caption{Change of the sidedness of jets with source and 
		core powers}
    \label{fig:sidedness}
\end{figure}

The general statistics of Table 1 is given in Table 2. Column 1 gives
the optical type of radio sources. The unidentified objects are listed as
U. Column 2 gives the source number. Columns 3, 5, 7 and 9 are the
minima and 4, 6, 8 and 10 the maxima of red-shift, total, jet, and core
powers at 1.4 GHz, 1.4 GHz and 5 GHz, respectively. The trend of the 
sidedness of jets varying with the core and total powers is 
given in Fig.~\ref{fig:sidedness}. All BL Lac objects, radio quasars
and high luminosity radio galaxies are one-sided. 

%
%%%--------------------------------------------------------------------

%%---------------------------------------------------------------------

\section{Discussion}

Jets are detected in the red-shift range from $ Z_{min} = 0.0012$ to $ 
Z_{max} = 3.886$. The closest radio source with detected jets is the
nearest active galaxy Cen A (1327+427) at 3.7 Mpc. Jets are detected
in QSOs in a red-shift range from 
$Z_{min} = 0.104$ to $ Z_{max} = 3.886$. The minimum value is very
close the lower limit of red-shift defining a QSO in Hewitt and
Burbidge (\cite{Hewitt}). The maximum of the red-shift dramatically 
increases from 2.877 in Paper I to 3.886. Because of the importance of 
high red-shift radio galaxies in the study of the formation and evolution 
of galaxies, many very powerful instruments have been dedicated to the
observations of radio galaxies with high redshift and consequently
brought about the detection of a large number of jets. The
highest red-shift for radio galaxies in Paper I is 0.574, which is
much smaller than 3.395 given in Table 2. The ranges of total,
jet, and core powers do not change a lot, and are comparable to those
in Paper I. 

Observations of compact steep-spectrum sources is another important
subject in the theory of galaxy formation. Only pc-scale jets are
detected in compact flat-spectrum and steep-spectrum sources.
In the galaxy formation theory, compact steep-spectrum sources are at
the early stages of the evolution of AGNs and the detected pc-scale
jets are instrinsically small. On the other hand, compact flat-spectrum 
sources are active galaxies with a small view angle of the jet to
the line of sight and the pc jets are the projections of the kpc jets.

We classify radio sources as radio galaxies, radio quasars, BL Lac 
objects, and Seyfert galaxies, based on the optical properties of the
sources. Jets are commonly detected in all types of sources with different 
detection rates. Large-scale jets are detected in a large fractions of 
weak radio galaxies and extended radio quasars, although the fraction
becomes lower in the lower range of power of core radio emission
(Parma et al. \cite{Parma}). This may result from the selection effect
as a consequence of the correlation between opening angle and radio
power (Bridle \cite{Bridle1}; Parma et al. \cite{Parma}; Paper I).
Jets are detected in a smaller fraction of distant radio galaxies
than in radio quasars with similar radio power and with instruments 
operating under similar parameters. 

We gave in this paper a new list of 661 extra-galactic radio sources with
detected jets. The list has two significant features: 
pc-scale jets are detected in many compact flat-spectrum and compact
steep-spectrum sources and kpc-scale jets are detected in a large
number of radio galaxies and radio quasars.

%
%______________________________________________________________

\begin{acknowledgements}

We are greatly indebated to Dr. A. Lanza for his helpful discussion.
We thank the referee Dr. R. Perley for his useful comments. 
FKL acknowledges support by the Swedish Natural Science Research
Council (NFR). He also thanks the director and staff of the Department
of Astronomy and Astrophysics, G\"oteborg University \& Chalmers 
University of Technology, for their warm hospitality. 

\end{acknowledgements}

%
%-----------------------------------------------------------------

%-----------------------------
\newpage
\onecolumn
%%%%%%%%%%%%%%%%	Table 1      %%%%%%%%%%
%Title: 		A new list of extra-galactic radio jets
%Reference number: 	A&A/2001/1596
%Authors:		F.K. Liu and Y.H. Zhang
%
\onecolumn
\begin{description}
	\item[Table 1.] 661 radio sources with detected jets
\end{description}
\nopagebreak
\def\hline{\noalign{\hrule}}
\tabskip=0pt
\offinterlineskip
%%%%%%%%%%%%%%%%%%%%%%%%%%%%%%%%%%%%%%%%%%%%%%%%%%%%%%%%%%%%%%%%%%%%%%%%%%%%%
\halign{\strut
\vrule\hskip3pt#\hfill&\vrule\hskip3pt#\hfill&\hfill#\hfill&\hfill#\hfill&
\hfill#\hfill&\hfill#\hfill&\hfill#\hfill&\hfill#\hfill
&\hfill#\hfill\vrule&\hskip3pt#\hfill&\vrule#\cr
\hline
\hfill\quad\ IAU\quad\ \hfill  &\hfill\quad NAME\quad\ \hfill &\quad ID \quad  &$\quad \lg P_j^{1.4}$\quad &\quad $\lg P_t^{1.4}$\quad &\quad $\lg P_c^5$\quad &\quad $d_j$\quad &\ SID\ &\
Mor\ &\hfill\quad  References\qquad\qquad\quad \hfil &\cr
         &        & Z    & W/Hz & W/Hz & W/Hz & Kpc &   &   &           &\cr
\hline
0003-066 &	  &  G	 &	&26.22 &26.22 &	    & 1 &   &\cite{k981151295}&\cr
W	 &	  &0.35  &	&      &      &0.034&   &   &		&\cr
\hline
0007+106 &IIIzW 2 &  G	 &	&25.33 &23.85 &	    & 1 &   &\cite{k981151295}&\cr
SW	 &	  &0.09  &	&      &      &0.019&   &   &		&\cr
\hline
0007+332 &4C33.01 &  Q	 &	&26.55 &24.57 &	    & 1 &   &\cite{gc91171}&\cr
S	 &	  &0.743 &26.13 &      &      &162  &   &   &		&\cr
\hline
0007+124 &4C12.03 &  G   &	& 25.41& 22.74&     & T &   &\cite{lp91537}&\cr
NE	 &	  &0.11	 &	&      &      &66   &   &   &		&\cr
SW(C-J)	 &	  &	 &	&      &      &     &   &   &		&\cr
\hline
0015-229 &	  &  G	 &	&27.42 &25.32 &     & 1 &   &\cite{cr1091}&\cr
SW	 &	  &2.010 &	&      &      &26   &   &   &		&\cr
\hline
0017+257 &4C25.01 &  Q   &      &26.62 &26.23 &     & 1 & C &\cite{gc91171,92}&\cr
 SW       &        &0.284 & 25.50&      &      &46   &   &   &		&\cr 
\hline
0017+154 &3C9     &  Q   &      &28.16 &25.45 &     & 1 & B &\cite{al94247,bh94766,29,143}&\cr
SE       &        &2.012 & 27.94&      &      &38   &   &   &\cite{gc91171,lb8763,24}&\cr 
NW(C-J)  &        &      & 25.98&      &      &6.1  &   &   &\cite{kd472115,240}&\cr
\hline
0017-207 &	  &  Q   &      &26.26 &23.67 &     & ? &   &\cite{i98300269}&\cr
NE       &        &0.545 & 	&      &      &58   &   &   &		&\cr 
SW(C-J)  &        &	 & 	&      &      &61   &   &   &		&\cr 
\hline
0019+230 &	  &  G	 &	&24.63 &22.67 &     & 1 &   &\cite{ol10841}&\cr
SE	 &	  &0.1338&	&      &      &26   &   &   &		&\cr
\hline
0022-297 &MRC	  &  Q	 &	&26.75 &25.45 &     & 1 &   &\cite{k98118327}&\cr
N	 &B2	  &0.406 &	&      &      &64   &   &   &		&\cr
\hline
0026+346 &S4	  &  G	 &	&      &      &     & 1 &   &\cite{fc11195}&\cr
NE	 &OB343	  &	 &	&      &      &     &   &   &		&\cr
\hline
0031+060 &	  &  G	 &	& 24.71&22.48 &     & 2 & C &\cite{ol10841}&\cr
N	 &	  &0.1334&	&      &      &17   &   &   &		&\cr
S	 &	  &	 &	&      &      &     &   &   &		&\cr
\hline
0031+403 &6C	  &  	 &	&      &      &     & 1 &   &\cite{b98295265}&\cr
NW	 &	  &	 &	&      &      &     &   &   &		&\cr
\hline
0033+183 &3C14    &  Q   &      &27.65 &25.58 &     & 1 &   &\cite{gc91171}&\cr
SE       &        &1.469 &27.32 &      &      &64   &   &   &\cite{29}&\cr
\hline
0033+079 &4C08.04 &  Q	 &	&27.08 &26.16 &     & 1 & C &\cite{lb8763}&\cr
SE	 &OB056	  &1.578 &	&      &      &11   &   &   &		&\cr
\hline
0034-014 &3C15	  &  G	 &	&25.41 &23.48 &     & 1 &   &\cite{lb29120,m98496203,m9912281,s00542667}&\cr
NW	 &	  &0.0730&24.59 &      &      &9.8  &   &   &		&\cr
SE(C-J)	 &	  &	 &22.87	&      &      &6.6  &   &   &		&\cr
\hline
0034+254 &B2      &  G   &      & 23.13&21.77 &     & 2 & C &\cite{cf91362}&\cr
E        &UGC367  &0.0321& 22.3 &      &      & 19  &   &   &\cite{67,175}&\cr
W        &        &      & 22.3 &      &      & 19  &   &   &		&\cr
\hline
0035+180 &	  &  G   &	&24.70 &23.27 &     & 2 & C &\cite{ol10841}&\cr
NW	 &	  &0.1448&	&      &      &94   &   &   &		&\cr
SW	 &	  &	 &	&      &      &87   &   &   &		&\cr
\hline
0035+121 &PKS	  &  Q   &	&27.67 &26.97 &     & 1 &   &\cite{b98293257}&\cr
N	 &B2	  &1.395 &	&      &      &9.4  &   &   &		&\cr
\hline
0038-019 &4C-02.04&  Q	 &	&27.63 &26.44 &     & 1 & L &\cite{lb8763}&\cr
S	 &PKS	  &1.690 &	&      &      &52   &   &   &		&\cr
\hline
0039+211 &4C21.05 &  G   &      & 24.89& 23.60&     & 2 & C &\cite{162,163}&\cr
S        &        &0.1017&23.09 &      &      &18   &   &   &		&\cr
NE       &        &      &23.09 &      &      &	    &   &   & 		&\cr
\hline
0040+517 &3C20	  &  G	 &	&26.52 &23.07 &     & 1 & S &\cite{ha288859}&\cr
NW	 &	  &0.174 &24.42	&      &      &47   &   &   &		&\cr
\hline
0043+201 &4C20.04 &  G   &	& 25.06&23.07 &     & 2 &   &\cite{167}&\cr
N	 &	  &0.1063&22.71 &      &      &39   &   &   &		&\cr
S	 &	  &   	 &22.70	&      &      &52   &   &   &		&\cr
\hline
0046+203 &	  &  	 &	&      &      &     & 1 & L &\cite{w98300790}&\cr
S	 &	  &	 &	&      &      &     &   &   &		&\cr
\hline
0046+226 &	  &  	 &	&      &      &     & 1 & L &\cite{w98300790}&\cr
N	 &	  &	 &	&      &      &     &   &   &		&\cr
\hline
0047+241 &	  &  G   &	&24.17 &23.08 &     & 2 & C &\cite{ol10841}&\cr
E	 &	  &0.0818&	&      &      &28   &   &   &		&\cr
SE	 &	  &      &	&      &      &28   &   &   &		&\cr
\hline
0048+509 &3C22	  &  Q	 &	&27.30 &24.98 &     & 1 &   &\cite{fb931690}&\cr
NW	 &	  &0.937 &	&      &      &55   &   &   &		&\cr
\hline
0050-220 &	  &  G   &	&23.57 &22.98 &     & 2 &   &\cite{ol10841}&\cr
NW	 &	  &0.0587&	&      &      &54   &   &   &		&\cr
SE	 &	  &	 &	&      &      &43   &   &   &		&\cr
\hline
0051+291 &4C29.01 &  Q   &	& 27.47& 26.90&	    & 1 & C &\cite{lb8763}&\cr
NE	 &	  &1.828 & 26.76&      &      &8.5  &   &   &\cite{7,239}&\cr
\hline
0053-016 &	  &  G   &      &24.34 &22.19 &     & 2 & C &\cite{162}	&\cr
NW	 &        &0.042 &22.54 &      &      &19   &   &   &	 	&\cr
SE	 &        &      &22.54 &      &      &19   &   &   &	 	&\cr
\hline
0053+260B&	  &  G	 &	&25.04 &23.00 &	    & 1 & C &\cite{98}	&\cr
NE	 &	  &0.1916&24.32 &      &      &23   &   &   &		&\cr
SE(C-J)	 &	  &	 &24.02	&      &      &47   &   &   &	 	&\cr
\hline
0055+300 &NGC315  &  S1  &      & 24.08& 23.24&     & T & L &\cite{km94729,cf91362,vg40881,c99519108}&\cr
SE(C-J)  &UGC597  &0.0167&      &      &      &240  &   &   &\cite{mk123423,25,29,49,26,67}&\cr
NW       &        &      &      &      &      &240  &   &   &\cite{81,135,112,259,k981151295}&\cr
\hline
0055-016 &3C29	  &  G	 &	&25.10 &23.32 &     & 2 &   &\cite{mk931023,x001202950}&\cr
SE	 &	  &0.045 &	&      &      &41   &   &   &		&\cr
NW	 &        &      &      &      &      &     &   &   &	 	&\cr
\hline
0055+265 &NGC326  &  G   &      & 24.61& 22.29&     & 2 & S &\cite{cf91362}&\cr
SE       &UGC601  &0.0472&      &      &      &27   &   &   &\cite{29,112}&\cr
NW       &        &      &      &      &      &27   &   &   &		&\cr
\hline
0104+321 &3C31    &  G   &      & 24.21& 22.45&     & 2 & C &\cite{cf91362,pc911960,lc474179,29,49}&\cr
NW       &NGC383  &0.0169& 23.48&      &      &14   &   &   &\cite{46,38,67,81,112}&\cr
SE       &UGC689  &      & 23.39&      &      &14   &   &   &\cite{117,237}&\cr
         &B2      &      &      &      &      &     &   &   &		&\cr
\hline
0106+013*&4C01.02 &  Q	 &	&28.06 &28.48 &     & 1 &   &\cite{kw901057,p951555,b003547}&\cr
S	 &	  &2.107 &27.42 &      &      &19   &   &   &		&\cr
\hline
0106+729 &3C33.1  &  S1  &      & 26.11& 23.76&     & 1 &   &\cite{29,208}&\cr
SW       &        &0.181 &24.81 &      &      &140  &   &   &		&\cr
\hline
0108-146 &PKS	  &  G	 &	&24.60 &23.80 &     & 1 &   &\cite{pc911960,ow80501}&\cr
SE	 &	  &0.1033&23.72 &      &      &1.4  &   &   &		&\cr
\hline
0109+200 &TEX	  &  Q	 &	&26.97 &25.98 &     & 1 &   &\cite{w98300790}&\cr
SW	 &	  &0.746 &      &      &      &11   &   &   &		&\cr
\hline
0110+297 &4C29.02 &  Q   &	&26.13 &25.04 &	    & 1 &   &\cite{gc91171}&\cr
NE	 &	  &0.363 &25.90 &      &      &140  &   &   &\cite{40,253}&\cr
\hline
0110+152 &	  &  G   &      & 24.34&22.06 &	    & 2 &   &\cite{167,ow80501}&\cr
NW	 &	  &0.0477&22.49 &      &      &20   &   &   &		&\cr
SE	 &        &    	 &22.28 &      &      &20   &   &   &		&\cr
\hline
0114-476 &	  &  G	 &	& 25.87&23.16 &     & 1 &   &\cite{ss96257}&\cr
NW	 &	  &0.146 &	&      &      &174  &   &   &		&\cr
SE(C-J)	 &	  &	 &	&      &      &200  &   &   &		&\cr
\hline
0116+082 &PKS	  &  S2	 &	&27.18 &25.54 &     & 1 & L &\cite{cv484193}&\cr
S	 &	  &0.594 &	&      &      &0.28 &   &   &		&\cr
\hline
0123-016A&NGC541  &  G   &	& 23.53& 21.79&	    & 2 & C &\cite{36,162,276,278}&\cr
NE	 &        &0.0182&	&      &      &22   &   &   &		&\cr
SW(C-J)	 &	  &	 &	&      &      &     &   &   &		&\cr
\hline
0123-016B&3C40    &  G   &      & 24.38& 22.02&     & 2 & B &\cite{mk931023}&\cr
NE       &NGC545  &0.0185&      &      &      &38   &   &   &\cite{29,31,162}&\cr
SW       &        &      &      &      &      &--   &   &   &		&\cr
\hline
0124+189 &	  &  G	 &	&24.42 &22.59 &     & 2 &   &\cite{ol10841}&\cr
N	 &	  &0.0420&	&      &      &12   &   &   &		&\cr
S	 &	  &	 &	&      &      &11   &   &   &		&\cr
\hline
0127+233 &3C43    &  Q   &	& 28.01&26.01 &     & 1 & L &\cite{ns91513,vbf9256,l98299467}&\cr
S	 &	  &1.46  & 27.63&    &      &5.1  &   &   &\cite{ss95629,ss91225}&\cr
\hline
0128+002 &A208	  &  G	 &	&24.82 &23.13 &     & 2 &   &\cite{bb9353}&\cr
NE	 &	  &0.1032&	&      &      &13   &   &   &		&\cr
SW	 &	  &	 &	&      &      &13   &   &   &		&\cr
\hline
0130+24  &4C24.02 &  Q   &      & 26.21& 25.11&     & 1 &   &\cite{gc91171}&\cr
E        &        &0.457 &25.85 &      &      &92   &   &   &\cite{29,253}&\cr
\hline
0133+207 &3C47	  &  Q	 &	& 26.89&25.23 &     & 1 &   &\cite{bh94766,fl9163}&\cr
SW	 &	  &0.425 &25.08 &      &      &129  &   &   &\cite{dtb289753}&\cr
C-J	 &	  &	 &$<$24.34&    &      &96   &   &   &		&\cr
\hline
0134+329 &3C48	  &  Q	 &	& 27.42&25.00 &     & 1 &   &\cite{wt352313,c00528201}&\cr
N	 &	  &0.367 &	&      &      &0.15 &   &   &		&\cr
\hline
0136+396 &4C39.04 &  G   &	& 25.76& 24.29&     & 1 &   &\cite{248}&\cr
SW	 &	  &0.211 &	&      &      &333  &   &   &		&\cr
\hline
0136+185 &	  &  G	 &	&23.92 &22.44 &     & 2 &   &\cite{ol10841}&\cr
NW	 &	  &0.0690&	&      &      &22   &   &   &		&\cr
SE	 &	  &	 &	&      &      &22   &   &   &		&\cr
\hline
0138+136 &3C49	  &  Q	 &	&27.14 &24.49 &	    & 1 &   &\cite{la94p471}&\cr
E	 &	  &0.62  &	&      &      &4.1  &   &   &		&\cr
\hline
0139+073A\ &	  &  G	 &	&23.26 &22.31 &     & 1 &   &\cite{ol10841}&\cr
E	 &	  &0.0616&	&      &      &37   &   &   &		&\cr
\hline
0141+061 &A245AB  &  G	 &	&24.04 &22.25 &     & 2 &   &\cite{ow80501}&\cr
NW	 &	  &0.089 &	&      &      &92   &   &   &		&\cr
SE	 &	  &	 &	&      &      &187  &   &   &		&\cr
\hline
0149+358 &NGC708  &  G   &      & 22.62& 21.31&     & 2 & S &\cite{bb91371}&\cr
W        &B2      &0.0160& 21.93&      &      &5.6  &   &   &\cite{29,175,176}&\cr
E        &        &      & 21.88&      &      &5.9  &   &   &		&\cr
\hline
0149+218 &	  &  Q   &      & 27.29& 27.55&     & 1 &   &\cite{k981151295}&\cr
N        &	  &1.32  & 	&      &      &0.024&   &   &		&\cr
\hline
0153+744 &S5	  &   Q	 &	&28.62 &27.81 &     & 1 & C &\cite{fc11195,hk324857}&\cr  
NE	 &	  &2.338 &	&      &      &0.066&   &   &		&\cr
\hline
0156-252 &	  &   G	 &	&27.45 &25.85 &     & 1 &   &\cite{cr1091}&\cr
NE	 &	  &2.090 &	&      &      &11   &   &   &		&\cr
\hline
0157+393B&4C39.05 &  G   &      & 24.34& 23.66&	    & ? & C &\cite{248} &\cr
SW	 &        &0.072 &	&      &      &66   &   &   &		&\cr
\hline
0202+091 &J0204+0903&    &      &      &      &     & 1 & C &\cite{p0053490}&\cr
E        &	  &	 &      &      &      &     &   &   &		&\cr
SW(C-J)  &        &      &      &      &      &     &   &   &		&\cr
\hline
0202+149*&	  &  Q	 &	&27.61 &27.23 &     & 1 & S &\cite{p00358451}&\cr
NW	 &	  &0.833 &	&      &      &0.029&   &   &		&\cr
\hline
0206+35  &UGC1651 &  G   &      & 24.52& 23.15&     & 2 & S &\cite{29,154,175,176}&\cr
NW       &4C35.03 &0.0375& 23.46&      &      &17   &   &   &		&\cr
SE       &        &      & 23.32&      &      &20   &   &   &		&\cr
\hline
0211-122 &	  &  G	 &	&27.30 &25.30 &     & 1 & L &\cite{cr1091}&\cr
SE	 &	  &2.336 &	&      &      &28   &   &   &		&\cr
\hline
0212+735 &S5	  &  Q   &	& 28.59& 28.20&     & 1 &   &\cite{pr88114,c9051,k981151295}&\cr
E	 &	  &2.367 &	&      &      &0.053&   &   &\cite{fc105299}&\cr
\hline
0219+421 &NGC891  &  G   &	& 21.68&20.38 &	    & 2 &   &\cite{cf91362}&\cr
SW	 &UGC1831 &0.002 &	&      &      &7.1  &   &   &\cite{248}	&\cr
NE	 &        & 	 &	&      &      &5.9  &   &   &		&\cr
\hline
0220+427 &3C66B   &  G   &      & 24.69& 22.59&     & 2 & C &\cite{js93128,fg91562,ha96273,84,85}&\cr
NE       &        &0.0215&23.90 &      &      &45   &   &   &\cite{ma91l55,fb284911,29,49,159}&\cr
SW       &        &      &23.90 &      &      &5.5  &   &   &\cite{117,119,248,272,x001202950}&\cr
         &        &      &      &      &      &     &   &   &\cite{t00317623,s00542667}&\cr
\hline
0221+067 &PKS	  &  Q	 &	&26.81 &26.25 &     & 1 &   &\cite{fc11195}&\cr
NW	 &	  &0.511 &	&      &      &0.029&   &   &		&\cr
\hline
0224+671 &4C67.05 &  Q   &	&      &      &	    & 1 &   &\cite{160}	&\cr
N	 &        &      &      &      &      &     &   &   &		&\cr
\hline
0229+341 &3C68.1  &  Q   &	& 27.65& 24.37&	    & 1 &   &\cite{bh94766}&\cr
N	 &	  &1.238 & 26.79&      &      &90   &   &   &\cite{24}	&\cr
S(C-J)	 &	  &   	 & 25.22&      &      &47   &   &   &		&\cr
\hline
0234+315 &3C68.2  &  G   &	& 27.08&23.72 &	    & 1 &   &\cite{b97292758}&\cr
SE	 &        &1.575 &      &      &      &46   &   &   &		&\cr
\hline
0238+085 &NGC1044 &  G   &      & 23.80& 22.54&     & 2 &   &\cite{vb43667}	&\cr
NW       &4C08.11 &0.0214&22.43 &      &      &43   &   &   &\cite{26}	&\cr
SE	 &	  & 	 &22.43 &      &      &     &   &   &		&\cr
\hline
0238+100 &MC	  &  Q   &      & 27.29& 26.16&	    & 1 & C &\cite{lb8763}&\cr
SE	 &	  &1.816 & 25.97&      &      &37   &   &   &\cite{7,143}&\cr
NW(C-J)	 &        &      &      &      &      &27   &   &   &		&\cr
\hline
0238-084 &NGC1052 &  S3	 &	&22.96 &22.43 &     & 2 &   &\cite{fc11195,c98500129,k981151295}&\cr
NE	 &PKS	  &0.005 &	&      &      &0.004&   &   &		&\cr
SW	 &	  &	 &	&      &      &0.001&   &   &		&\cr
\hline
0240-002 &3C71    &  S1  &      & 22.94& 20.99&     & 2 & S &\cite{gb464198,gb458136,b001202904}&\cr
NE       &NGC1068 &0.004 & 21.81&      &      &0.35 &   &   &\cite{29,58,185,262,263}&\cr
SW       &        &      & 21.83&      &      &0.25 &   &   &		&\cr
\hline
0244+152 &UGC2252 &  G	 &	&      &      &     & 2 &   &\cite{cf91362}&\cr
NW	 &	  &	 &	&      &      &     &   &   &		&\cr
SE	 &	  &	 &	&      &      &     &   &   &		&\cr
\hline
0247+467 &	  &  G   &	& 23.97&22.56 &     & 2 & C &\cite{248}	&\cr
NW	 &	  &0.029 &	&      &      &191  &   &   &		&\cr 
NE	 &        &      &	&      &      &175  &   &   &		&\cr
\hline
0253+633 &	  &	 &	&      &      &	    & 1 & L &\cite{pb92655}&\cr
SE	 &	  &	 &	&      &      &	    &   &   &		&\cr
\hline
0255+05  &3C75A,B &  G   &      & 24.61& 22.40&     & 2 & C &\cite{29,166,171,173,187}&\cr
E        &        &0.0241&      &      &      &30   &   &   &		&\cr
W        &        &      &      &      &      &30   &   &   &		&\cr
\hline
0256+132 &4C13.17B&  G   &      & 24.07& 22.30&     & 2 & C &\cite{29,162}&\cr
NW       &        &0.0748& 22.92&      &      &15   &   &   &		&\cr
NE	 &        & 	 & 22.92&      &      &14   &   &   &		&\cr
\hline
0256+366 &UGC2456 &  S2  &      & 22.18& 21.26&     & 2 &   &\cite{n99120209}&\cr
NW       &MRK1066 &0.012 & 	&      &      &0.25 &   &   &		&\cr
SE	 &        & 	 & 	&      &      &0.31 &   &   &		&\cr
\hline
0258+350 &4C34.09 &  G   &	&23.88 &22.48 &     & 2 &   &\cite{ss95629}&\cr
NW	 &	  &0.020 &	&      &      &0.33 &   &   &		&\cr
SE	 &	  &	 &	&      &      &0.20 &   &   &		&\cr
\hline
0300+162 &3C76.1  &  G	 &	& 24.52&21.94 &     & 2 &   &\cite{lp91537}&\cr
NW	 &PGC11499&0.0316&	&      &      &18   &   &   &		&\cr
SE	 &	  &	 &	&      &      &22   &   &   &		&\cr
\hline
0301+336 &	  &  	 &	&      &      &	    & 1 &   &\cite{w98300790}&\cr
E	 &        &      &	&      &      &     &   &   &		&\cr
\hline
0304+575 &GT	  &  G	 &	&      &      &	    & 1 &   &\cite{tg95238}&\cr
NE	 &        &      &	&      &      &     &   &   &\cite{63,64}&\cr
SW	 &	  &	 &	&      &      &     &   &   &		&\cr
\hline
0305+039 &3C78    &  G   &      & 24.83& 23.77&     & 2 & S &\cite{29,213,m9912281,s00542667}&\cr
NE	 &NGC1218 &0.0289&      &      &      &0.6  &   &   &		&\cr
SW(C-J)  &        &      &	&      &      &     &   &   &		&\cr
\hline
0307+444 &4C44.07 &  Q   &	& 27.40& 26.61&	    & 1 &   &\cite{211} &\cr
	 &        &1.165 & 26.12&      &      &7.3  &   &   &		&\cr
\hline
0309+390 &4C39.11 &  S1  &	&25.72 &24.31 &     & 1 &   &\cite{xr99297}&\cr
SW	 &	  &0.161 &	&      &      &48   &   &   &		&\cr
\hline
0313+683 &WNB     &  G   &	&24.92 &23.56 &     & 1 &   &\cite{s98336455}&\cr
SW	 &	  &0.0901&	&      &      &733  &   &   &		&\cr
\hline
0314+416 &NGC1265 &  G   &      & 24.76& 22.15&     & 2 & C &\cite{29,53,164,165,166}&\cr
E        &3C83.1B &0.0255& 22.91&      &      &  18 &   &   &\cite{x991172626,d0012933}&\cr
W        &        &      & 22.91&      &      &16   &   &   &		&\cr
\hline
0316+308 &	  &  	 &	&      &      &	    & 1 &   &\cite{w98300790}&\cr
SE	 &        &      &	&      &      &     &   &   &		&\cr
\hline
0316+413\%&NGC1275&  BL  &      & 24.66& 24.85&     & 1 &   &\cite{kw93375,pg90477,vr411552,r98131451,h991181942}&\cr
SE       &3C84    &0.0177& 23.00&      &      &5.0  &   &   &\cite{vr430l41,wr430l45,184,198}&\cr
N(C-J)   &MRK1505 &      & 22.70&      &      &5.0  &   &   &\cite{5,29,143,266,209}&\cr
         &Perseus A&     &      &      &      &     &   &   &\cite{d98498111,w00530233}&\cr
\hline
0317-023 &4C-02.15&  Q	 &	&27.68 &27.09 &     & 1 & C &\cite{b003547}&\cr
E	 &	  &2.092 &      &      &      &8.7  &   &   &		&\cr
\hline	
0319-454 &	  &  G	 &	&25.26 &22.82 &     & 2 &   &\cite{ss9437,jm80137}&\cr
NE	 &	  &0.0633&23.10 &      &      &190  &   &   &		&\cr
SW(C-J)  &	  &	 &22.69	&      &      &295  &   &   &		&\cr
\hline	
0320-37  &For A   &  G   &      & 24.73& 21.15&     & T &   &\cite{29,68}&\cr
SE	 &NGC1316 &0.0063& 21.95&      &      &3.6  &   &   &		&\cr
NW	 &	  &	 & 21.65&      &      &3.6  &   &   &		&\cr
\hline
0325+395 &	  &  	 &	&      &      &     & 1 & C &\cite{pb92655}&\cr
S	 &	  &	 &	&      &      &     &   &   &		&\cr
\hline
0326+396 &VV7.08.14& G   &      & 24.06& 22.70&     & 2 & S &\cite{cf91362,bb91371}&\cr
E        &UGC2775 &0.0243&22.88 &      &      &  41 &   &   &\cite{29,67}&\cr
W        &B2      &      &23.22 &      &      &  41 &   &   &		&\cr
\hline
0327+246B&	  &  G   &	& 23.71& 22.71&     & 2 &   &\cite{162}	&\cr
W	 &	  &0.1063&	&      &      &19   &   &   &		&\cr
E	 &	  &	 &	&      &      &19   &   &   &		&\cr
\hline
0328-033 &MRK612  &  S2  &	& 21.55& 20.18&     & 2 & S &\cite{n99120209}&\cr
NE	 &	  &0.020 &	&      &      &0.39 &   &   &		&\cr
W	 &	  &	 &	&      &      &2.2  &   &   &		&\cr
\hline
0331-013 &3C89	  &  G	 &	&25.77 &23.96 &     & 2 & C &\cite{21}	&\cr
NW	 &	  &0.1386&	&      &      &52   &   &   &		&\cr
SE	 &	  &	 &	&      &      &45   &   &   &		&\cr
\hline
0333+321 &NRAO 140&  Q   &	& 28.36&27.23 &	    & 1 &   &\cite{150,fc96543,k981151295,l0054166}&\cr
SE	 &	  &1.258 &	&      &      &0.042&   &   &\cite{149}&\cr
\hline
0335+096 &	  &  G   &	& 23.70& 22.36&     & 2 & C &\cite{162}	&\cr
NW	 &	  &0.035 & 22.49&      &      &23   &   &   &		&\cr
SW	 &	  &      &22.49	&      &      &23   &   &   &		&\cr
\hline
0336-35  &PKS     &  G   &      & 22.82& 20.41&     & 2 &   &\cite{29,68}&\cr
S        &NGC1399 &0.0049& 21.81&      &      &  8.1&   &   &		&\cr
N        &        &      & 21.81&      &      &  8.1&   &   &		&\cr
\hline
0356+102 &3C98	  &  S2	 &	&25.02 &21.81 &     & 1 &   &\cite{21,lb29120}&\cr
NE	 &	  &0.0306&23.45	&      &      & 60  &   &   &		&\cr
SW(C-J)	 &	  &	 &	&      &      &     &   &   &		&\cr
\hline
0400+258 &PKS	  &  Q	 &	&28.53 &27.51 &     & 1 & C &\cite{fc11195}&\cr
NE	 &	  &2.109 &	&      &      &0.075&   &   &		&\cr
\hline
0404+768 &	  &  G	 &	& 27.42&25.82 &     & 1 &   &\cite{df9527,tr46395}&\cr
W	 &	  &0.5985&	&      &      &0.25 &   &   &		&\cr
\hline
0405-123 &PKS	  &  G   &	& 27.39& 26.60&	    & 1 &   &\cite{209}	&\cr
N	 &OF-109  &0.574 &	&      &      & 40  &   &   &		&\cr
\hline
0406-244 &	  &  G	 &	&27.64 &25.41 &     & 1 &   &\cite{cr1091}&\cr
NW	 &	  &2.440 &	&      &      &17   &   &   &		&\cr
\hline
0409+229 &3C108	  &  Q	 &	&27.21 &27.06 &	    & 1 &   &\cite{sj90408}&\cr
NW	 &4C22.08 &1.215 &	&      &      &11   &   &   &		&\cr
\hline
0410+110 &3C109	  &  S1	 &	&26.80 &25.15 &     & 1 &   &\cite{gf435116}&\cr
SE	 &	  &0.3066&24.74	&      &      &136  &   &   &		&\cr
\hline
0411+141 &4C14.11 &  G	 &	&26.51 &24.17 &     & 1 &   &\cite{ha288859}&\cr
SE	 &	  &0.206 &23.48	&      &      &9.5  &   &   &		&\cr
\hline
0411+055 &4C05.19 &  Q	 &	&28.43 &27.17 &     & 1 & C &\cite{r00362845}&\cr
NE	 &PKS, MG &2.639 &27.14	&      &      &0.35 &   &   &		&\cr
NW(C-J)	 &DA 131  &      &26.49	&      &      &0.12 &   &   &		&\cr
\hline
0414+009 &1H	  &  BL  &	& 25.30&24.70 &     & 1 &   &\cite{lmk93875}&\cr
NE	 &	  &0.287 &	&      &      & 20  &   &   &		&\cr
\hline
0415+379 &3C111   &  S1  &      & 26.29& 24.47&     & 1 &   &\cite{29,91,99,132,135}&\cr
NE       &        &0.0485& 24.20&      &      & 77  &   &   &\cite{138,139,lb29120}&\cr
\hline
0417-181 &	  &  G	 &	&27.63 &24.72 &     & 1 &   &\cite{cr1091}&\cr
S	 &	  &2.773 &	&      &      &10   &   &   &		&\cr
\hline
0420-014*&PKS	  &  Q	 &	&27.82 &27.26 &     & 1 & S &\cite{b0036065}&\cr
SW	 &	  &0.915 &	&      &      &0.052&   &   &		&\cr
\hline
0422+148 &J0424+1442&    &      &      &      &     & 1 &   &\cite{p0053490}&\cr
W        &	  &	 &      &      &      &     &   &   &		&\cr
\hline
0429+415 &3C119	  &  Q	 &	&28.11 &26.26 &     & 1 &   &\cite{n99344402}&\cr
SW	 &	  &1.023 &	&      &      &0.12 &   &   &		&\cr
\hline
0430+052*&3C120   &  S1  &      & 24.76& 24.93&	    & 1 &   &\cite{mw9154,3,6,13,29,252}&\cr
W        &        &0.0334& 24.07&      &      &100  &   &   &\cite{35,109,187,249,250}&\cr
	 &        &  	 &	&      &      &	    &   &   &\cite{251,265,g98499221,b991221,k981151295}&\cr
	 &        &  	 &	&      &      &	    &   &   &\cite{g002892317,h99518213,g9952129}&\cr
\hline
0434+299 &	  &  	 &	&      &      &	    & 1 &   &\cite{w98300790}&\cr
SW	 &        &      &	&      &      &     &   &   &		&\cr
\hline
0437-244 &MRC	  &  Q	 &	&26.61 &25.02 &     & 1 &   &\cite{k98118327}&\cr
S	 &B2	  &0.84  &	&      &      &256  &   &   &		&\cr
\hline
0439+083 &	  &  G	 &	&25.07 &23.12 &     & 1 &   &\cite{ol10841}&\cr
SE	 &	  &0.1517&	&      &      &129  &   &   &		&\cr
\hline
0445+097 &4C09.17 &  Q   &	& 27.78&27.29 &     &1  &   &\cite{7,lb8763,l9912411}&\cr
SW	 &	  &2.110 &	&      &      & 4.9 &   &   &		&\cr
\hline
0445+44  &3C129   &  G   &      & 24.58& 21.92&     & T & S &\cite{29,111,281,l00313617}&\cr
N        &4U      &0.0208& 23.10&      &      &  8.8&   &   &		&\cr
S        &        &      & 23.14&      &      &  8.8&   &   &		&\cr 
\hline
0448+520 &3C130   &  G   &	& 26.79&23.57 &	    & 2 &   &\cite{14,112,113,h98298569,h99349381}&\cr
SW	 &	  &0.1090&23.17	&      &      &132  &   &   &		&\cr
NE	 &        &  	 &23.32	&      &      &119  &   &   &		&\cr
\hline
0449-175 &PKS     &  G   &      & 23.97& 22.03&     & 2 &   &\cite{29,68}&\cr
SE       &        &0.0313& 20.99&      &      &17   &   &   &		&\cr
NW	 &        &      & 20.99&      &      & 10  &   &   &		&\cr
\hline
0454-220 &PKS	  &  Q	 &	&26.89 &25.74 &    & 1 &    &\cite{ae932054,i98300269}&\cr
SE	 &	  &0.53	 &	&      &      & 200&   &    &		&\cr
NW(C-J)	 &	  &	 &	&      &      & 69 &   &    &		&\cr
\hline
0457+052 &A526	  &  G	 &	&24.48 &22.97 &     & 2 &   &\cite{ow80501}&\cr
NW	 &	  &0.098 &	&      &      &47   &   &   &		&\cr
SE	 &	  &	 &	&      &      &36   &   &   &		&\cr
\hline
0457+054 &	  &  G   &	&23.71&$<$21.81&    & 2 & C &\cite{162}	&\cr
E	 &	  &0.0541&22.92	&      &      &51   &   &   &		&\cr
NW	 &	  &      &22.92	&      &      &--   &   &   &		&\cr
\hline
0459+252 &3C133   &  G   &      & 26.72& 25.33&     & 1 &   &\cite{29,206,220}&\cr
NE       &        &0.2775& 24.68&      &      &15   &   &   &		&\cr
\hline
0501+380 &3C134   &  G   &	&      &      &	    & 2 &   &\cite{132}	&\cr
N	 &	  &	 &	&      &      &	    &   &   &		&\cr 
S	 &	  &	 &	&      &      &	    &   &   &		&\cr
\hline
0504+030 &4C03.10 &  Q   &	&28.11 &27.54 &	    & 1 &   &\cite{b003547}&\cr
NW	 &	  &2.453 &	&      &      &7.5  &   &   &		&\cr 
SE(C-J)	 &	  &	 &	&      &      &8.0  &   &   &		&\cr
\hline
0507+179 &PKS	  &   	 &	&      &      &     & 1 &   &\cite{fc11195}&\cr
SW	 &	  &	 &	&      &      &     &   &   &		&\cr
\hline
0511+008 &3C135	  &  S1	 &	&25.69 &22.41 &     & 1 &   &\cite{lb29120}&\cr
SW	 &	  &0.1273&23.26	&      &      &52   &   &   &		&\cr
\hline
0514+109 &B2      &      &      &      &      &     & 1 &   &\cite{b98293257}&\cr
SW       &        &      &      &      &      &     &   &   &		&\cr
\hline
0514-16  &PKS     &  Q   &      & 27.25& 27.32&     & 1 &   &\cite{29}	&\cr
         &        &1.278 &      &      &      & 33  &   &   &		&\cr
\hline
0518-458 &PKS     &  G   &      & 25.93& 24.01&     & 1 &   &\cite{p9732812,s99123447,t001191695}	&\cr
W	 &Pictor A&0.0342&23.63 &      &      & 176 &   &   &		&\cr
\hline
0518+165 &PKS	  &  Q   &	& 27.79& 26.40&     & 1 &   &\cite{ff90333,ns91513,as93752}&\cr
NE	 &4C16.12 &0.759 & 27.27&      &      & 3.2 &   &   &\cite{dc95671,as91215,l98299467}&\cr
SW(C-J)	 &3C138   &      &	&      &      &     &   &   &\cite{70,71,72,cd325493}&\cr
	 &OG130.2 &      &	&      &      &     &   &   &		&\cr
\hline
0521+281 &3C139.2 &  G   &	&      &      &	    & 2 &   &\cite{132}	&\cr
NE	 &        &      &	&      &      &     &   &   &		&\cr
SW	 &        &      &	&      &      &     &   &   &		&\cr
\hline
0521-365 &PKS     &  BL  &      & 25.83& 24.75&	    & 1 &   &\cite{ma91369,sm90l41,t98500673,k981151295,s99526643}&\cr
SE	 &        &0.061 & 24.87&      &      &11   &   &   &\cite{51,68,117,118}&\cr
\hline
0528+134*&PKS	  &  Q	 &	& 28.62&27.97 &     & 1 &   &\cite{zm43291,b99341418}&\cr
NE	 &	  &2.07	 &	&      &      &0.020&   &   &		&\cr
\hline
0537+531 &	  &  Q   &	&27.22 &27.21 &     & 1 &   &\cite{tv95345}&\cr
NW	 &	  &1.275 &	&      &      &0.15 &   &   &		&\cr
\hline
0538+474 &	  &	 &	&      &      &	    & 1 &   &\cite{pb92655}&\cr
W	 &	  &	 &	&      &      &	    &   &   &		&\cr
\hline
0538+498*&3C147   &  Q   &      & 27.95& 26.78&     & 1 &   &\cite{vbf9256,as90362,za91650,l98299467,n00357891}&\cr
SW       &        &0.545 & 27.63&      &      & 0.86&   &   &\cite{sr90140,pw981,29,195,198}&\cr
	 &	  &	 &	&      &      &     &   &   &\cite{199,200,233,257}&\cr
\hline
0545+265 &	  &	 &	&      &      &     & 1 &   &\cite{275}	&\cr
NE	 &	  &	 &	&      &      &     &   &   &		&\cr
\hline
0546-329 &PKS     &  G   &      & 25.47& 23.90&     & 1 &   &\cite{29,68,jm80137}&\cr
SW       &        &0.147 & 24.76&      &      &  200&   &   &		&\cr
\hline
0548-322 &PKS	  &  BL  &	& 24.39&23.60 &     & 1 &   &\cite{lmk93875}&\cr
E	 &	  &0.069 &	&      &      &  6  &   &   &		&\cr
\hline
0548+165 &	  &  Q   &	& 26.85&25.50 &     & 1 & L &\cite{m9833210}&\cr
W	 &	  &0.474 & 26.80&      &      & 0.64&   &   &		&\cr
\hline
0549-074 &NGC2110 &  S1  &	& 22.18&21.63 &     & 2 & S &\cite{n99120209,m00529816}&\cr
NE	 &MCG-1-15-4&0.007&20.54&      &      & 0.16&   &   &		&\cr
SW	 &	  &      & 20.54&      &      & 0.14&   &   &		&\cr
\hline
0553-205 &MC	  &  Q	 &	&26.81 &26.26 &     & 1 &   &\cite{lb8763}&\cr
NW	 &	  &1.544 &	&      &      &9.8  &   &   &		&\cr
\hline
0605-085 &OH-010  &  Q   &	& 27.86& 27.37&	    & 1 & B &\cite{160} &\cr
E	 &	  &0.87  &	&      &      &21   &   &   &		&\cr
\hline
0609+710 &MRK3	  &  S1  &	& 23.54& 21.85&	    & 2 & S &\cite{kg93893,k99518117}&\cr
E	 &	  &0.0137&22.36	&      &      &0.5  &   &   &\cite{186}	&\cr
W	 &	  &	 &22.14 &      &      &0.2  &   &   &		&\cr
\hline
0637-752*&PKS	  &  Q   &	&27.84 &27.40 &     & 1 & L &\cite{t98497594,s0054069,c00542655}&\cr
W	 &	  &0.654 &      &      &      &73   &   &   &		&\cr
\hline
0645+744 &MRK6	  &  S1  &	&23.03 &20.81 &     & 1 &   &\cite{kh961283,g9833313}&\cr
SE	 &IC450	  &0.0193&22.61 &      &      &0.18 &   &   &		&\cr
\hline
0651+542 &3C171	  &  S2	 &	&26.35 & 23.13&     & 1 &   &\cite{ha288859}&\cr
NW 	 &	  &0.2384&24.50	&      &      &11   &   &   &		&\cr
SE(C-J)	 &	  &	 &24.11	&      &      &10   &   &   &		&\cr
\hline
0658+329 &A567B	  &  G	 &	& 24.78&22.87 &     & 2 &   &\cite{ow80501}&\cr
NW	 &	  &0.127 &	&      &      &47   &   &   &		&\cr
SW	 &	  &	 &	&      &      &47   &   &   &		&\cr
\hline
0658+330 &B2      &  G   &      & 24.98& 23.68&     & 2 & C &\cite{ow80501}&\cr
E        &A567A   &0.127 &      &      &      &  55 &   &   &\cite{29,162}&\cr
W        &        &      &	&      &      &     &   &   &		&\cr
\hline
0702+749 &3C173.1 &  G	 &	&26.41 &24.00 &     & 1 &   &\cite{ha288859}&\cr
N	 &	  &0.292 &23.86	&      &      &69   &   &   &		&\cr
\hline
0704+351A&4C35.16A&  G   &      & 24.28& 21.83&     & 2 & L &\cite{26,162}&\cr
N        &        &0.078 &      &      &      &  17 &   &   &		&\cr
SE       &        &      &      &      &      &  17 &   &   &		&\cr
\hline
0707+476 &B3	  &  Q	 &	&27.81 &27.30 &     & 1 & C &\cite{fc11195}&\cr
NE	 &S4	  &1.31	 &	&      &      &0.037&   &   &		&\cr
\hline
0710+118 &3C175   &  Q   &	& 27.28& 25.16&     & 1 & L &\cite{bh94766,dtb289753,h9951184}&\cr
SW	 &	  &0.768 & 25.67&      &      & 112 &   &   &\cite{23,24,bh9491}&\cr
C-J	 &	  &	 &$<$23.77&    &      & 77  &   &   &		&\cr
\hline
0710+439 &S4	  &  S	 &	&27.15 &24.82 &     & 1 &   &\cite{tr46395}&\cr
N	 &	  &0.518 &	&      &      &0.041&   &   &		&\cr
S(C-J)	 &        &      &	&      &      &     &   &   &		&\cr
\hline
0711+356 &OI318	  &  Q	 &	&27.99 &26.14 &     & 1 &   &\cite{mp425568}&\cr
NW	 &	  &1.626 &	&      &      &0.09 &   &   &		&\cr
\hline
0712+534 &4C53.16 &  G   &      & 24.83& 22.96&     & 1 &   &\cite{gc91171}&\cr
NW       &        &0.064 & 24.54&      &      &  13 &   &   &\cite{29,46}&\cr
\hline
0714+457 &	  &  Q	 &	& 26.72&26.74 &     & 1 & C &\cite{hb1001}&\cr
SE	 &	  &0.940 &	&      &      &0.14 &   &   &		&\cr
\hline
0714-292 &M4-1	  &  S2	 &	& 21.34&19.28 &     & 2 &   &\cite{f98502199}&\cr
NW	 &ESO428-G14&0.0054&	&      &      &0.077&   &   &		&\cr
SE	 &	  &      &	&      &      &0.25 &   &   &		&\cr
\hline
0716+714 &S5	  &  BL  &	&      &      &	    & 1 &   &\cite{g00313627}&\cr
NW	 &	  &	 &	&      &      &     &   &   &		&\cr
\hline
0720+381 &	  &  G   &	& 25.24& 24.92&	    & 1 &   &\cite{248}&\cr
NE	 &	  &0.220 &	&      &      &  230&   &   &		&\cr
\hline
0720+670 &	  &  G	 &	&23.85 &23.58 &     & 2 & C &\cite{ol10841,ow80501}&\cr
NW	 &	  &0.0864&	&      &      &29   &   &   &		&\cr
SE	 &	  &	 &	&      &      &29   &   &   &		&\cr
\hline
0723-008 &PKS	  &  BL	 &	&25.99 &24.89 &     & 1 & C &\cite{bp96415,fc11195}&\cr
NW	 &	  &0.13	 &	&      &      &0.072&   &   &		&\cr
\hline
0723+679*&3C179   &  Q   &      & 27.39& 26.62&     & 1 &   &\cite{a92l61,29,34,35,109,170}&\cr
W        &        &0.846 &26.21 &      &      &59   &   &   &\cite{174,194,195,198,s98298877}&\cr
\hline
0730+257 &PKS     &  Q   &	& 27.73& 26.37&	    & 1 & B &\cite{lb8763}&\cr
NW	 &	  &2.686 & 27.59&      &      & 8.2 &   &   &\cite{143}	&\cr
\hline
0733+597 &	  &  G   &	&23.99 &22.93 &     & 2 &   &\cite{tv95345}&\cr
N	 &	  &0.040 &	&      &      &0.018&   &   &		&\cr
S	 &	  &	 &	&      &      &0.006&   &   &		&\cr
\hline
0735+178*&PKS     &  BL  &	&$>$26.58&$>$26.62&  & 1& S &\cite{gw435128,g99519642}&\cr
S	 &        &$>$0.424&	&      &      &$>$6.6&  &   &\cite{4,160,195}&\cr
\hline
0738+441 &	  &  G	 &	&24.92 &23.86 &     & 2 &   &\cite{ol10841,ow80501}&\cr
N	 &	  &0.1170&	&      &      &39   &   &   &		&\cr
S	 &	  &	 &	&      &      &32   &   &   &		&\cr
\hline
0740+828 &	  &  Q	 &	& 28.10&27.34 &     & 1 &   &\cite{xr99297}&\cr
N	 &	  &1.991 &	&      &      &0.090&   &   &		&\cr
\hline
0740+380 &3C186	  &  Q	 &	&27.34 &25.40 &	    & 1 & S &\cite{ss91225,l98299467}&\cr
NW	 &	  &1.06	 &26.82	&      &      &5.3  &   &   &		&\cr
\hline
0742+318 &4C31.30 &  Q   &      & 26.55& 26.24&	    & 1 &   &\cite{mo912026}&\cr
NW       &        &0.462 &      &      &      & 210 &   &   &\cite{29,92}&\cr
\hline
0745+521 &	  &  G	 &	& 24.42&22.68 &     & 2 &   &\cite{ol10841}&\cr
E	 &	  &0.0678&	&      &      &73   &   &   &		&\cr
W	 &	  &	 &	&      &      &64   &   &   &		&\cr
\hline
0745+241 &	  &  Q	 &	& 26.56&25.88 &     & 1 &   &\cite{k981151295}&\cr
NW	 &	  &0.41  &	&      &      &0.033&   &   &		&\cr
\hline
0748+126 &PKS	  &  Q	 &	&27.26 &27.21 &     & 1 &   &\cite{mb93298}&\cr
SE	 &	  &0.889 &	&      &      &61   &   &   &		&\cr
\hline
0751+298 &4C29.27 &  Q	 &	& 27.47&26.80 &     & 1 & C &\cite{lb8763}&\cr
NW	 &B2	  &2.106 &	&      &      &9.6  &   &   &		&\cr
\hline
0752+258 &B2      &  Q   &	& 26.03& 24.81&	    & 1 &   &\cite{2,215}&\cr
SE	 &OI287	  &0.446 & 	&      &      & 18  &   &   &		&\cr
NW(C-J)  &        &      &	&      &      & 25  &   &   &		&\cr
\hline
0755+379 &NGC2484 &  G   &      & 24.49& 23.59&     & 1 &   &\cite{xr99297,c00362871,b0031411}&\cr
SE	 &B2      &0.0413& 23.52&      &      &  28 &   &   &\cite{67,175,248}&\cr
NW(C-J)	 &3C189	  &	 &	&      &      &     &   &   &		&\cr
\hline
0756+272 &A610AB  &  G	 &	&24.66 &22.68 &     & 2 &   &\cite{ow80501,ol10841}&\cr
NE	 &	  &0.0956&	&      &      &53   &   &   &		&\cr
SW	 &	  &	 &	&      &      &39   &   &   &		&\cr
\hline
0757+395 &	  &  G   &	& 23.60& 23.16&	    & 2 &   &\cite{248}&\cr
NW	 & 	  &0.057 & 	&      &      & 76  &   &   &		&\cr
SE	 &	  &      &	&      &      & 70  &   &   &		&\cr
\hline
0758+143 &3C190	  &  Q	 &	&27.78 &25.95 &	    & 1 & B &\cite{ss91225,l98299467}&\cr
SW	 &	  &1.2	 &27.19	&      &      &7.1  &   &   &		&\cr
\hline
0800+608 &OJ061   &  Q   &	& 26.82&25.35 &	    & 1 &   &\cite{jb90750}&\cr
NE	 &	  &0.689 & 26.33&      &      & 70  &   &   &\cite{229}&\cr
\hline
0802+103 &3C191	  &  Q   &	& 28.12& 26.36&	    & 1 & C &\cite{al94247,gc91171,lb8763}&\cr
SE	 &	  &1.956 &27.67 &      &      & 12  &   &   &\cite{7,50,181}&\cr
\hline
0805+046 &4C05.34 &  Q   &      & 27.90& 27.62&	    & 1 &   &\cite{7,74}&\cr
NW	 &	  &2.877 &	&      &      &17   &   &   &		&\cr
\hline
0806-103 &3C195	  &  S2	 &	&25.73 &23.88 &     & 1 &   &\cite{mk931023}&\cr
SW	 &	  &0.110 &	&      &      &87   &   &   &		&\cr
\hline
0812+367 &B2      &  Q   &      & 27.23& 27.10&     & 1 & C &\cite{29,192}&\cr
NW       &        &1.025 &      &      &      &  30 &   &   &		&\cr
\hline
0812+02  &4C02.23 &  Q   &      & 26.68& 25.60&     & 1 &   &\cite{pg86365}&\cr
NW       &        &0.402 &      &      &      &32   &   &   &\cite{29,89,107,208}&\cr
\hline
0820+225 &PKS	  &  BL  &      &27.47 & 26.42&     & 1 &   &\cite{g003191109}&\cr
SW       &        &0.951 &27.44 &      &      &0.15 &   &   &		&\cr
\hline
0821+695 &8C	  &  G	 &	& 25.70&24.36 &     & 2 &   &\cite{lr93721,l0035663}&\cr
N	 &	  &0.538 &	&      &      &11   &   &   &		&\cr
S	 &	  &	 &	&      &      &12   &   &   &		&\cr
\hline
0821+621 &4CP62.12B&  Q	 &	&26.42 &26.31 &     & 1 &   &\cite{lr93721,tv10737}&\cr
SW	 &	  &0.542 &	&      &      &88   &   &   &		&\cr
\hline
0824+294 &3C200   &  G   &	& 26.73& 24.96&     & 1 &   &\cite{gc91171,bh9491}&\cr
SE	 &	  &0.458 & 25.72&      &      & 96  &   &   &\cite{24,29,43}&\cr
\hline
0826+180 &B2      &  BL   &	& 24.61& 23.96&     & 1 &   &\cite{b98293257}&\cr
SE	 &TEX	  &0.089  & 	&      &      & 1.1 &   &   &		&\cr
\hline
0828+193 &	  &  G	 &	&26.97 &25.68 &     & 1 &   &\cite{cr1091}&\cr
NE	 &	  &2.572 &	&      &      &26   &   &   &		&\cr
\hline
0828+493 &OJ448	  &  BL	 &	&26.73 &25.90 &     & 1 & C &\cite{g99307725}&\cr
NE	 &	  &0.548 &25.86	&      &      &0.084&   &   &		&\cr
\hline
0829+046 &PKS	  &  BL	 &	&25.55 &25.35 &     & 1 & L &\cite{l98504702}&\cr
NE	 &	  & 0.18 &	&      &      &59 &   &   &		&\cr
\hline
0829+280 &	  &  	 &	&      &      &	    & 1 &   &\cite{w98300790}&\cr
SE	 &        &      &	&      &      &     &   &   &		&\cr
\hline
0831+557 &4C55.16 &  G	 &	& 26.97&25.45 &     & 1 & C &\cite{pw981,fc11195}&\cr
NW	 &	  &0.2404&	&      &      &0.40 &   &   &		&\cr
\hline
0833+654 &3C204   &  Q   &      & 27.39& 25.69&     & 1 &   &\cite{bh94766,dtb289753,h9951184}&\cr
NW       &        &1.112 & 27.29&      &      &  52 &   &   &\cite{24,29,43,174}&\cr
C-J	 &	  &	 &$<$23.90&    &      &60   &   &   &		&\cr
\hline
0833+585 &        &  Q   &      & 27.52& 27.64&     & 1 & L &\cite{mb93298,pw981}&\cr
SE       &        &2.101 &      &      &      &  45 &   &   &\cite{29}	&\cr
\hline
0835+580 &3C205   &  Q   &      & 27.86& 25.73&     & 1 &   &\cite{l98115895,h9951184}&\cr
SW       &B2      &1.536 &      &      &      &  37 &   &   &		&\cr
\hline
0836+195 &4C19.31 &  Q	 &	&27.24 &26.56 &     & 1 & S &\cite{lb8763}&\cr
NE	 &OJ160	  &1.691 &	&      &      &57   &   &   &		&\cr
\hline
0836+710*&S5	  &  Q   &	& 28.51& 27.67&	    & 1 & L &\cite{kh90271,264,l9834060,o98334489}&\cr
SW	 &4C71.07 &2.16  &	&      &      &30   &   &   &		&\cr
\hline
0836+299A&4C29.30 &  S2  &      & 24.51& 22.97&     & 1 &   &\cite{vc454735}&\cr
SW       &        &0.0643& 23.33&      &      &  22 &   &   &\cite{41,175,280}&\cr
\hline
0836+290 &B2      &  G   &      & 24.73& 23.86&     & 1 & L &\cite{cf95643,167,ow80501}&\cr
N	 &OJ261   &0.0650& 23.85&      &      &92   &   &   &\cite{vc454735}&\cr
S(C-J)	 &	  &	 &	&      &      &62   &   &   &\cite{167,175,276}&\cr
\hline
0838+133 &3C207   &  Q   &	& 27.23& 26.45&     & 1 & L &\cite{gc91171,29,108}&\cr
E	 &	  &0.684 &26.93	&      &      & 25  &   &   &\cite{bh9491}&\cr
\hline
0838+325 &	  &  G	 &	&24.51 &24.13 &     & 1 &   &\cite{ol10841}&\cr
S	 &	  &0.0694&	&      &      &95   &   &   &		&\cr
\hline
0839+616 &4C61.19 &  Q	 &	&27.13 &25.06 &     & 1 &   &\cite{h9951184}&\cr
NW	 &	  &0.862 &	&      &      &32   &   &   &		&\cr
\hline
0839+186 &DW	  &  Q	 &	&27.39 &27.23 &     & 1 &   &\cite{p0053490}&\cr
NE	 &	  &1.272 &	&      &      &0.070&   &   &		&\cr
\hline
0844+540 &4C54.17 &  G	 &	&24.56 &22.59 &     & 2 &   &\cite{xr99297}&\cr
SE	 &	  &0.0453&	&      &      &14   &   &   &		&\cr
NW	 &	  &	 &	&      &      &9.5  &   &   &		&\cr
\hline
0844+319 &4C31.32 &  G   &      & 24.88& 23.35&     & 1 & S &\cite{cf95643,29,148,175}&\cr
N        &        &0.0675& 23.35&      &      & 65  &   &   &		&\cr
SE	 &	  &	 &	&      &      &     &   &   &		&\cr
\hline
0850+140 &3C208   &  Q   &      & 27.52& 26.27&	    & 1 &   &\cite{bh94766,24,29,43}&\cr
SW       &        &1.110 & 26.17&      &      &  22 &   &   &\cite{dtb289753}&\cr
C-J	 &	  &      & 25.33&      &      &22   &   &   &		&\cr
\hline
0850+581 &4C58.17 &  Q   &	& 27.35& 27.26&	    & 1 &   &\cite{gc91171,12,34,107,183}&\cr
SE	 &	  &1.322 &26.54	&      &      & 38  &   &   &\cite{228,k981151295}	&\cr
\hline
0855+143 &3C212   &  Q   &      & 27.61& 25.38&     & 1 &   &\cite{29}	&\cr
NW       &        &1.049 &      &      &      &  22 &   &   &		&\cr
\hline
0859+681 &	  &  Q   &	&27.33 &27.24 &     & 1 &   &\cite{tv95345}&\cr
NE	 &	  &1.499 &	&      &      &0.12 &   &   &		&\cr
\hline
0859+470 &4C47.29 &  Q	 &	&28.17 &27.27 &     & 1 & B &\cite{fc11195}&\cr
N	 &	  &1.462 &	&      &      &0.24 &   &   &		&\cr
\hline
0859-140 &	  &  Q	 &	&27.99 &27.36 &     & 1 &   &\cite{k981151295}&\cr
SE	 &	  &1.33  &	&      &      &0.087&   &   &		&\cr
\hline
0902+343 &	  &  G	 &	&27.94 &26.33 &     & 1 &   &\cite{co94480,lgl95939}&\cr
NW	 &	  &3.395 &27.84	&      &      &4.0  &   &   &		&\cr
\hline
0903+169 &3C215	  &  Q   &	& 26.27& 24.63&     & 1 & S &\cite{bh94766,gc91171,24,h9951184}&\cr
SE	 &	  &0.411 & 25.30&      &      & 40  &   &   &		&\cr
NW(C-J)	 &	  &      & 24.30&      &      & 70  &   &   &		&\cr
\hline
0905-097 &26W20	  &  G   &	& 24.15& 23.16&     & 1 & L &\cite{s98335443}&\cr
SE	 &	  &0.054 & 23.45&      &      & 305 &   &   &		&\cr
\hline
0905+420 &B2	  &  Q   &	&      &      &     & 1 &   &\cite{a982991159}&\cr
SW	 &	  &	 & 	&      &      &     &   &   &		&\cr
\hline
0906+430*&3C216   &  Q   & 	& 27.37&26.86 &	    & 1 & L &\cite{fp92459,tg95522,pw981,p0052983,v991181931}&\cr
SE	 &  	  &0.668 &26.39	&      &      &	0.6 &   &   &\cite{ap96p83,11,183}&\cr
\hline
0907-091 &	  &  G   &	& 24.24& 23.26&     & T & C &\cite{162}	&\cr
NW	 &        &0.1375&	&      &      & 26  &   &   &		&\cr
SE	 &	  & 	 &	&      &      & 17  &   &   &		&\cr
\hline
0908-103 &	  &  G   &	& 25.06&23.39 &	    & 2 & C &\cite{ow80501,167}&\cr
NW	 &	  &0.129 &	&      &      & 195 &   &   &		&\cr
SE	 &	  & 	 &	&      &      &	195 &   &   &		&\cr
\hline
0908+376 &B2      &  G   &      & 24.89& 23.50&     & T & C &\cite{29,148,175}&\cr
NE       &        &0.1047& 24.18&      &      &  29 &   &   &		&\cr
SW       &        &      & 23.71&      &      &  23 &   &   &		&\cr
\hline
0909+162 &	  &  G	 &	&24.17 &21.99 &     & 2 &   &\cite{ol10841,bb9353}&\cr
E	 &	  &0.0851&	&      &      &19   &   &   &		&\cr
SW	 &	  &	 &	&      &      &19   &   &   &		&\cr
\hline
0910+411 &IRAS	  &  G	 &	&24.51 &23.60 &     & 2 &   &\cite{hw41582}&\cr
NW	 &	  &0.442 &	&      &      &29   &   &   &		&\cr
SE	 &	  &	 &	&      &      &34   &   &   &		&\cr
\hline
0912+211 &	  &  	 &	&      &      &	    & 1 &   &\cite{w98300790}&\cr
NE	 &        &      &	&      &      &     &   &   &		&\cr
\hline
0913+38  &B2      &  G   &      & 24.27&$<$22.48&   & T &   &\cite{175}	&\cr
W        &        &0.0711& 23.39&      &      &  16 &   &   &		&\cr
E        &        &      & 23.06&      &      &$>$8 &   &   &		&\cr
\hline
0915+32  &B2      &  G   &      & 24.0 & 22.56&     & 2 & S &\cite{67,175}&\cr
N        &        &0.0620& 22.90&      &      &  51 &   &   &		&\cr
S        &        &      & 22.90&      &      &  56 &   &   &		&\cr
\hline
0915-118 &3C218   &  S3  &  	& 26.14&23.95 &	    & 1 & S &\cite{tp9041,24,21,t96p133}&\cr
NE	 &Hydra A & 0.053&	&      &      & 5   &   &   &		&\cr
SW(C-J)  &        &      &      &      &      & 5   &   &   &		&\cr
\hline
0917+458 &3C219   &  S1  &      & 26.45& 24.18&	    & 1 &   &\cite{cb385173,nr99349,b9934257}&\cr
SW       &        &0.1744& 24.87&      &      &39   &   &   &\cite{24,28,190,211,242}&\cr
NE(C-J)	 &        &      & 23.22&      &      &3.3  &   &   &		&\cr
\hline
0917+624*&	  &  Q	 &	&27.62 &27.66 &     & 1 &   &\cite{sq9627}&\cr
NW	 &	  &1.446 &	&      &      &0.14 &   &   &		&\cr
\hline
0919+218 &4C21.25 &  Q	 &	&27.24 &25.85 &	    & 1 &   &\cite{sj90408}&\cr
NE	 &	  &1.421 &26.37 &      &      &22   &   &   &		&\cr
\hline
0920+313 &B2	  &  Q	 &	&25.94 &26.27 &	    & 1 &   &\cite{w98300790}&\cr
SE	 &        &0.892 &	&      &      &12   &   &   &		&\cr
\hline
0923+392*&4C39.25 &  Q	 &	&27.37 & 27.73&	    & 1 &   &\cite{c9051,ak9393,jb93128,a97327513,l00364391}&\cr
NE	 &        &0.699 & 26.09&      &      &  8.0&   &   &\cite{kw901057,gm952586,147,k981151295,a00361529}&\cr
	 &	  &	 &	&      &      &	    &   &   &\cite{mz91491,am402160,151,l0054166}&\cr
\hline
0925+420 &B	  &  G	 &	&      &23.86 &     & 1 &   &\cite{s00315371}&\cr
NE	 &	  &0.365 &	&      &      &248  &   &   &		&\cr
\hline
0927+586 &	  &   Q	 &	&      &      &	    & 1 &   &\cite{pb92655}&\cr
N	 &	  &	 &	&      &      &	    &   &   &		&\cr
\hline
0936-041 &A841AB  &  G	 &	&24.70 &23.16 &     & 2 &   &\cite{ow80501}&\cr
NE	 &	  &0.094 &	&      &      &56   &   &   &		&\cr
SW	 &	  &	 &	&      &      &75   &   &   &		&\cr
\hline
0937+391 &4C39.27 &  Q	 &	&26.55 &24.71 &     & 1 &   &\cite{gc91171,pg86365,bh9491}&\cr
SE	 &	  &0.617 &26.28	&      &      &82   &   &   &		&\cr
\hline
0938+39  &4C39.27 &  Q   &      & 26.99& 25.00&     & 1 &   &\cite{29,253}&\cr
E        &        &0.618 &      &      &      &  96 &   &   &		&\cr
\hline
0941+261 &OK270	  &  Q	 &	&28.12 &27.39 &     & 1 & C &\cite{lb8763}&\cr
NE	 &B2	  &2.910 &	&      &      &7.1  &   &   &		&\cr
\hline
0941+522 &OL568	  &  Q	 &	&26.57 &26.17 &     & 1 &   &\cite{hb1001}&\cr
S	 &	  &0.565 &	&      &      &0.13 &   &   &		&\cr
\hline
0953+254 &OK290	  &  Q	 &	&26.66 &26.53 &     & 1 &   &\cite{mb93298,l0054166}&\cr
SW	 &	  &0.712 &	&      &      &50   &   &   &		&\cr
\hline
0954+658*&S4	  &  BL	 &	&26.28 &25.48 &     & 1 & L &\cite{g00315229}&\cr
SW	 &	  &0.386 &	&      &      &18   &   &   &		&\cr
\hline
0954+443 &7C	  &  Q	 &	&      &      &     & 1 &   &\cite{r99307293}&\cr
NW	 &	  &      &	&      &      &     &   &   &		&\cr
\hline
0955+428 &7C	  &  Q	 &	&      &      &     & 1 &   &\cite{r99307293}&\cr
SE	 &	  &      &	&      &      &     &   &   &		&\cr
\hline
0957+003 &4C00.34 &  Q   &      & 27.03& 26.02&     & 1 &   &\cite{gc91171,29,107,h921311}&\cr
W        &        &0.907 &26.65 &      &      &  76 &   &   &		&\cr
\hline
0957+561 &Double  &  Q   &      & 27.15& 25.96&     & 1 &   &\cite{cl952566,lh92453,29,b99520479}&\cr
NE       &QSO     &1.405 & 26.14&      &      &  22 &   &   &\cite{39,95,96,196,203}&\cr
\hline
0958+290 &3C234	  &  S1	 &	&26.30 &24.14 &     & 1 &   &\cite{ha288859}&\cr
NE	 &	  &0.1848&23.58	&      &      &128  &   &   &		&\cr
\hline
1000+263 &	  &  	 &	&      &      &	    & 1 &   &\cite{w98300790}&\cr
S	 &        &      &	&      &      &     &   &   &		&\cr
\hline
1001+226 &4C22.26 &  Q   &      & 26.94& 25.73&	    & 1 &   &\cite{gc91171,29,253}&\cr
N	 &        &0.974 &27.04 &      &      &  31 &   &   &		&\cr
\hline
1003+351 &3C236   &  S3  &      & 25.78& 24.64&     & 1 &   &\cite{mk123423,7,26,29}&\cr
SE       &        &0.0989&      &      &      &  0.4&   &   &		&\cr
\hline
1004+146 &NGC3121 &  G   &      &24.07 & 22.97&     & 2 &   &\cite{29,j200705}&\cr
NW       &        &0.031 &      &      &      &77   &   &   &		&\cr
SE	 &	  &	 &	&      &      &77   &   &   &		&\cr
\hline
1004+130 &4C13.41 &  Q   &      &25.91 & 23.87&     & 1 &   &\cite{29,74}&\cr
E 	 &        &0.240 &	&      &      &60   &   &   &		&\cr
W(C-J)	 &	  &	 &	&      &      &     &   &   &		&\cr
\hline
1005+28  &	  &  G   &      &24.25 & 22.6 &	    & 2 &   &\cite{175}	&\cr
N	 &        &0.1476& 23.28&      &      &40   &   &   &		&\cr
S	 &   	  &	 & 23.67&      &      &89   &   &   &		&\cr
\hline
1005+449 &3C237	  &  S   &      &27.92 & 24.65&	    & 1 &   &\cite{l98299467}&\cr
E	 &        & 0.88 & 	&      &      &0.70 &   &   &		&\cr
\hline
1007+417 &4C41.21 &  Q   &      &26.91 & 25.85&	    & 1 &   &\cite{29,170,174,210}&\cr
S	 &        &0.613 &    	&      &      &70   &   &   &		&\cr
\hline
1009+748 &4C74.16 &  G	 &	&27.27 &24.31 &	    & 1 &   &\cite{gc91171}&\cr
N	 &	  &0.81	 &26.93	&      &      &96   &   &   &		&\cr
\hline
1010+350 &	  &  G	 &	&27.14 &26.87 &     & 1 &   &\cite{hb1001}&\cr
E	 &	  &1.414 &	&      &      &0.054&   &   &		&\cr
\hline
1010+287 &B2	  &  	 &	&      &      &     & 1 &   &\cite{a982991159}&\cr
SE	 &	  &	 &	&      &      &     &   &   &		&\cr
\hline
1010+420 &7C	  &  Q	 &	&      &      &     & 1 &   &\cite{r99307293}&\cr
SW	 &	  &      &	&      &      &     &   &   &		&\cr
\hline
1011+496 &B2	  &  BL	 &	&25.25 &24.91 &	    & 1 &   &\cite{a982991159}&\cr
W	 &	  & 0.2  &	&      &      &1.6  &   &   &		&\cr
\hline
1012+232 &4C23.24 &  Q	 &	&26.45 &26.45 &	    & 1 &   &\cite{212}	&\cr
N	 &PKS	  &0.565 &	&      &      &41   &   &   &		&\cr
\hline
1015+359 &OL326   &  Q   &	&27.18 &27.16 &     & 1 &   &\cite{xr99297,k981151295}&\cr
SE	 &	  &1.226 &	&      &      &27   &   &   &		&\cr
\hline
1018+456 &7C      &  Q   &	&25.34 &24.89 &     & 1 &   &\cite{r99307293}&\cr
S	 &B3	  &0.364 &	&      &      &47   &   &   &		&\cr
\hline
1020+400 &4C40.25 &  Q	 &	&27.51 &26.70 &     & 1 &   &\cite{pw981,fc11195}&\cr
NW	 &B2	  &1.254 &	&      &      &1.5  &   &   &		&\cr
\hline
1022+194 &4C19.34 &  Q	 &	&27.32 &26.60 &     & 1 &   &\cite{fc11195}&\cr
NE	 &	  &0.828 &	&      &      &0.38 &   &   &		&\cr
\hline
1023+067 &3C243	  &  Q   &	&27.38 &26.47 &     & 1 &   &\cite{lb8763}&\cr
SE	 &	  &1.699 &	&      &      &34   &   &   &		&\cr
NW(C-J)	 &	  &	 &	&      &      &30   &   &   &		&\cr
\hline
1024+419 &	  &  Q	 &	&      &      &     & 1 &   &\cite{r99307293}&\cr
S	 &	  &      &	&      &      &     &   &   &		&\cr
\hline
1028+313 &B2	  &  Q	 &	&24.80 & 24.52&	    & 1 &   &\cite{74,92}&\cr
NW	 &OL347   &0.177 & 	&      &      &18   &   &   &		&\cr
\hline
1029+570 &HB 13   &  G   &      &23.81 &22.50 &	    & 2 &   &\cite{29,112,152,p98298637}&\cr
NW	 &        & 0.034&23.31 &      &      &210  &   &   &		&\cr
SE	 &        &      &-----	&      &      &167  &   &   &		&\cr
\hline
1030+602 &MRK34   &  S2	 &	&22.67 &21.35 &	    & 2 &   &\cite{f98502199}&\cr
NW	 &	  &0.0505&	&      &      &0.63 &   &   &		&\cr
SE	 &	  &	 &	&      &      &0.83 &   &   &		&\cr
\hline
1032+343 &TEX	  &  Q	 &	&25.87 &25.45 &     & 1 & C &\cite{r99307293}&\cr
S	 &	  &0.680 &	&      &      &124  &   &   &		&\cr
\hline
1033+003 &PKS     &  G	 &	&      &      &	    & 1 &   &\cite{29}	&\cr
SW	 &	  &	 &	&      &      &     &   &   &		&\cr
NE(C-J)	 &	  &	 &	&      &      &     &   &   &		&\cr
\hline
1034+461A&	  &  	 &	&      &      &     & 1 &   &\cite{r99307293}&\cr
NE	 &	  &      &	&      &      &     &   &   &		&\cr
\hline
1035+500 &	  &  Q	 &	&      &      &     & 1 &   &\cite{r99307293}&\cr
S	 &	  &      &	&      &      &     &   &   &		&\cr
\hline
1035+486 &	  &  Q	 &	&      &      &     & 1 & C &\cite{r99307293}&\cr
E	 &	  &      &	&      &      &     &   &   &		&\cr
W(C-J)	 &	  &      &	&      &      &     &   &   &		&\cr
\hline
1037+426 &	  &  Q	 &	&      &      &     & 1 &   &\cite{r99307293}&\cr
SW	 &	  &      &	&      &      &     &   &   &		&\cr
\hline
1038+528 &OL564	  &  Q	 &	&26.44 &26.39 &     & 1 &   &\cite{pg86365,tv10737}&\cr
NE	 &	  &0.677 &	&      &      &78   &   &   &		&\cr
\hline
1040+31  &4C29.41 &  G   &      &24.03 &22.77 &	    & 1 &   &\cite{175,176}&\cr
S	 &B2      &0.0360& 22.60&      &      &7.7  &   &   &		&\cr
\hline
1040+123*&3C245   &  Q	 &	& 27.80& 27.16&	    & 1 &   &\cite{fb9017,al94247,sj90408}&\cr
W	 &4C12.37 & 1.029&27.08 &      &      &19   &   &   &\cite{lp92545,34,79,110,rs114511}&\cr
\hline
1055+201 &4C20.24 &  Q	 &	&27.65 &27.01 &     & 1 &   &\cite{gc91171,mb93298,k981151295}&\cr
N	 &PKS	  &1.111 &27.54	&      &      &92   &   &   &		&\cr
\hline
1055+018 &4C01.28 &  Q	 & 	& 27.51& 27.51&	    & 1 & S &\cite{37,156,160,a9951887}&\cr
S	 &PKS, DA293&0.888&	&      &      &  125&   &   &\cite{mb93298}&\cr
	 &OL093   &      & 	&      &      &     &   &   &		&\cr
\hline
1058+726 &4C72.16 &  Q	 &	&26.43 &25.79 &	    & 1 &   &\cite{pb92655}&\cr
N	 &S5	  &0.375 &	&      &      &8.5  &   &   &		&\cr
\hline
1100+772 &3C249.1 &  Q   &      & 26.41& 25.00&     & 1 & C &\cite{ks941163,bh94766,mr9293}&\cr
E        &        & 0.311&25.35 &      &      &22   &   &   &\cite{24,29,43,145}&\cr
NW(C-J)	 &        &      &24.36 &      &      &10   &   &   &\cite{dtb289753}&\cr
\hline
1100+350 &	  &  Q	 &	&      &      &     & 1 &   &\cite{lgl95939}&\cr
E	 &	  &	 &	&      &      &     &   &   &		&\cr
\hline
1100+358 &	  &  G   &	& 27.06& 25.35&	    & 1 & S &\cite{b99303616}&\cr 
NE	 &	  & 1.44 &26.00	&      &      &29   &   &   &		&\cr
SW(C-J)	 &	  &      &--	&      &      &42   &   &   &		&\cr
\hline
1101+384 &MRK421  &  BL  &	& 23.68& 23.47&	    & 1 &   &\cite{4,288,k981151295,p99525176}&\cr 
NW	 &	  &0.0308&23.45	&      &      &0.003&   &   &		&\cr
\hline
1101-325 &PKS	  &  G	 &	&26.28 &25.82 &     & 1 &   &\cite{dw28485}&\cr
S	 &	  &0.3554&	&      &      &89   &   &   &		&\cr
\hline
1102+30  &B2	  &  G	 &	& 24.29& 22.76&     & 1 &   &\cite{175,284,285}&\cr
NE	 &        &0.0720&$>$22.84&    &      &$>$12&   &   &		&\cr
SW(C-J)	 &	  &	 &	&      &      &     &   &   &		&\cr
\hline
1103-006 &	  &  Q	 &	& 26.39&25.32 &     & 1 & S &\cite{ks941163,mr9293}&\cr
NW	 &	  &0.426 &	&      &      &43   &   &   &		&\cr
\hline
1104+167 &MC	  &  Q   &	& 26.61&26.30 &	    & 1 & C &\cite{h921311,107}&\cr
N	 &	  &0.632 &	&      &      &90   &   &   &		&\cr
\hline
1107-372 &NGC3557 &  G	 &	& 22.80&21.42 &     & 2 &   &\cite{19}	&\cr
E	 &	  &0.00952&	&      &      &15   &   &   &		&\cr
W	 &	  &	 &	&      &      &     &   &   &		&\cr
\hline
1107+483 &S4	  &  Q	 &	&26.70 &25.73 &     & 1 & L &\cite{p0053490}&\cr
S	 &	  &0.74  &	&      &      &0.12 &   &   &		&\cr
\hline
1108+27  &B2      &  G   &      &23.01 & 22.21&     & 1 &   &\cite{67,175,176}&\cr
W	 &        &0.0331& 22.80&      &      &19   &   &   &		&\cr
E	 &        &      & 21.92&      &      &12   &   &   &		&\cr
\hline
1108+411 &4C41.23 &  G	 &	&24.73 &22.91 &     & 2 & C &\cite{162}	&\cr
N	 &	  &0.0737&23.87	&      &      &5.1  &   &   &		&\cr
S	 &	  &	 &23.87	&      &      &5.1  &   &   &		&\cr
\hline
1113-178 &	  &  G	 &	&27.39 &25.66 &     & 1 &   &\cite{cr1091}&\cr
NW	 &	  &2.239 &	&      &      &9.3  &   &   &		&\cr
\hline
1113+295C&4C29.41 &  G   &      &24.67 & 22.94&	    & 1 &   &\cite{ol10841,154,175,176}&\cr
W	 &B2      &0.0489& 23.34&      &      &22   &   &   &		&\cr
E(C-J)	 &	  &	 &	&      &      &     &   &   &		&\cr
\hline
1116+28  & B2     &  G   &      &24.33 &23.17 &	    & 2 &   &\cite{175}	&\cr
E	 &        &0.0667& 23.20&      &      &50   &   &   &		&\cr
W   	 &    	  &	 & 23.70&      &      &75   &   &   &		&\cr
\hline
1117+543 &	  &   Q	 &	&      &      &	    & 1 &   &\cite{pb92655}&\cr
W	 &	  &	 &	&      &      &	    &   &   &		&\cr
\hline
1118+000 &PKS	  &  G	 &	&      &      &     & 2 &   &\cite{pc911960}&\cr
S	 &	  &	 &	&      &      &     &   &   &		&\cr
W	 &	  &	 &	&      &      &     &   &   &		&\cr
\hline
1120+013 &	  &  G	 &	&24.12 & 22.65&     & 2 & L &\cite{ol10841}&\cr
N	 &	  &0.0719&	&      &      &19   &   &   &		&\cr
S	 &	  &	 &	&      &      &29   &   &   &		&\cr
\hline
1121+809 &S5	  &  BL  &	&      &      &     & 1 &   &\cite{fc11195}&\cr
N	 &	  &	 &	&      &      &     &   &   &		&\cr
\hline
1122+390 &NGC3665 &  G   &      &21.76 &20.46 &	    & 2 &   &\cite{19,29,175,176}&\cr
W	 &B2      &0.0067& 21.17&      &      &4.0  &   &   &		&\cr
E	 &        &      & 21.25&      &      &3.4  &   &   &		&\cr
\hline
1125+37	 &4C37.30 &  Q	 &	&      &      &	    & 1 & L &\cite{na92647}&\cr
NE	 &	  &	 &	&      &      &	    &   &   &		&\cr
\hline
1130-037 &A1308B  &  G	 &	&24.54 &23.85 &     & 2 & C &\cite{ow80501}&\cr
N	 &	  &0.063 &	&      &      &168  &   &   &		&\cr
S	 &	  &	 &	&      &      &144  &   &   &		&\cr
\hline
1131+493 &IC708   &  G   &      &24.13 &22.74 &	    & 2 & C &\cite{ol10841}&\cr
NW       &        &0.0321&     	&      &      &44   &   &   &\cite{29,247}&\cr
SW	 &        &      &      &      &      &44   &   &   &		&\cr
\hline
1132+492 &IC711	  &  G   &      &23.73 &22.59 &	    & 2 & C &\cite{162} &\cr
NW	 &	  &0.0324& 22.01&      &      &13   &   &   &		&\cr
SW	 &	  &	 &22.01 &      &      &13   &   &   &		&\cr
\hline
1136-135 &PKS	  &  Q	 &	& 27.15& 26.27&	    & 1 &   &\cite{ae932054,211}&\cr
SE	 &	  & 0.554&	&      &      &38   &   &   &		&\cr
\hline
1137+18  &NGC3801 &  G   &      &23.06 &20.59 &	    & 2 & C &\cite{29}	&\cr
SE	 &        &0.0105&      &      &      &2.1  &   &   &		&\cr
SW	 &	  &	 &	&      &      &2.1  &	&   &		&\cr
\hline
1137+660*&3C263   &  Q	 &	&27.90 &25.98 &	    & 1 &   &\cite{bh94766,hz464715}&\cr
SE	 &    	  & 0.646& 25.54&      &      &63   &   &   &\cite{24,110,271,dtb289753}&\cr
C-J	 &        &      &$<$23.93&    &      &94   &   &   &		&\cr
\hline
1138-262 &PKS	  &  G	 &	&27.73 &25.49 &     & 1 &   &\cite{cr1091}&\cr
W	 &	  &2.156 &	&      &      &34   &   &   &		&\cr
\hline
1141+354 &4C35.26 &  G	 &	&27.15 &24.73 &     & 1 &   &\cite{lgl95939}&\cr
NE	 &	  &1.781 &	&      &      &12   &   &   &		&\cr
\hline
1142+198 &3C264,B2&  S	 &	&24.48 &23.09 &     & 1 & S &\cite{pc911960,c96p201,m9912281,l99513197,x001202950}&\cr
NE	 &NGC3862 &0.0208&23.67 &      &      &10   &   &   &\cite{cp402l37,lc474179,s00542667}&\cr
SW(C-J)  &UGC6723 &	 &	&      &      &     &   &   &\cite{cp402l37,21}&\cr
\hline
1142+052 &PKS     &  Q	 &	&27.50 &26.94 &     & 1 &   &\cite{b98293257}&\cr
NE	 &B2	  &1.342 &	&      &      &7.8  &   &   &		&\cr
\hline
1144+352 &B2	  & BL	 &	&24.39 &24.37 &	    & 1 & S &\cite{175,176,s9833370,s9934144,g99522101}&\cr
SE	 &Zw186.48&0.0630&22.80 &      &      &21   &   &   &		&\cr
NW(C-J)	 &	  &      &22.15 &      &      &     &   &   &		&\cr
\hline
1146+596 &NGC3894 &  G	 &	& 23.71&21.93 &     & 2 &   &\cite{tv95345,t98498619}&\cr
SE	 &UGC06779&0.01085&22.55&      &      &0.0019&  &   &		&\cr
NW	 &	  &	 &22.74	&      &      &0.0009&  &   &		&\cr
\hline
}
%%%---------------------------------------------------------------------------
\newpage
\begin{description}
	\item[Table 1] \it{continued}
\end{description}
\halign{\strut
\vrule\hskip3pt#\hfill&\vrule\hskip3pt#\hfill&\hfill#\hfill&\hfill#\hfill&
\hfill#\hfill&\hfill#\hfill&\hfill#\hfill&\hfill#
\hfill&\hfill#\hfill&\vrule\hskip3pt#\hfill&\vrule#\cr
\hline
\hfill\quad\ IAU\quad\ \hfill  &\hfill\quad NAME\quad\ \hfill &\quad ID \quad  &$\quad \lg P_j^{1.4}$\quad &\quad $\lg P_t^{1.4}$\quad &\quad $\lg P_c^5$\quad &\quad $d_j$\quad &\ SID\ &\
Mor\ &\hfill\quad  References\qquad\qquad\quad \hfill &\cr
         &        & Z    & W/Hz & W/Hz & W/Hz & Kpc &   &   &           &\cr
\hline
1147+245 &OM280	  &  BL  &	&      &      &	    & 1 & S &\cite{gs961877,g99307725}&\cr
W	 &B2,MG	  &	 &	&      &      &	    &   &   &		&\cr
	 &GC	  &	 &	&      &      &	    &   &   &		&\cr
\hline
1150+497 &4C49.22 &  Q	 &	& 26.43& 25.85&     & 1 & B &\cite{ag91644,gc91171,pw981,s98298877}&\cr
SW	 &	  & 0.334&26.14	&      &      &22   &   &   &\cite{37,170,174}&\cr
\hline
1150+812 &S5	  &  Q	 &	&27.76 &27.26 &     & 1 &   &\cite{fc11195}&\cr
SE	 &	  &1.25  &	&      &      &0.035&   &   &		&\cr
\hline
1155+266 &	  &  G	 &	&25.10 &22.87 &     & 2 &   &\cite{ol10841}&\cr
NW	 &	  &0.1120&	&      &      &66   &   &   &		&\cr
SE	 &	  &	 &	&      &      &56   &   &   &		&\cr
\hline
1155+186 &B2	  &  	 &	&      &      &     & 1 &   &\cite{b98293257}&\cr
SW	 &	  &	 &	&      &      &     &   &   &		&\cr
\hline
1156+295*&4C29.45 &  BL  &	&27.10 &26.99 &	    & 1 &   &\cite{mm90305,w98300790}&\cr
N	 &	  &0.729 &	&      &      &16   &   &   &		&\cr
\hline
1158+122 &MC2	  &  Q	 &	&27.14 &26.13 &     & 1 &   &\cite{lb8763}&\cr
SW	 &	  &2.018 &	&      &      &52   &   &   &		&\cr
\hline
1159+583 &4C58.23 &  G	 &	& 25.07& 22.81&     & 1 & C &\cite{167,ol10841}&\cr
NE	 &	  &0.1018&23.55	&      &      &14   &   &   &		&\cr
SW(C-J)	 &	  &	 &	&      &      &     &   &   &		&\cr
\hline
1201+205 &NGC4061 &  S2	 &	&23.58 &21.38 &     & 2 & C &\cite{vb43667}&\cr
E	 &	  &0.0236&21.75	&      &      &30   &   &   &		&\cr
W	 &	  &	 &21.75	&      &      &30   &   &   &		&\cr
\hline
1201+282 &	  &  G	 &	&24.68 &$<$21.89&   & ? &   &\cite{ol10841}&\cr
S	 &	  &0.1390&	&      &      &48   &   &   &		&\cr
\hline
1203+043 &	  &  	 &	&      &      &	    & 1 &   &\cite{mj92353}&\cr
SW	 &	  &	 &	&      &      &	    &   &   &		&\cr
\hline
1206+439 &3C268.4 &  Q	 &	& 27.66& 26.08&	    & 1 &   &\cite{lp92545,43}&\cr
SW	 &	  & 1.400& 25.90&      &      &21   &   &   &		&\cr
\hline
1208+396 &NGC4151 &  S1	 &	& 21.71&20.47 &     & 2 &   &\cite{pk93471,mp95355,u98496196,p983001071}&\cr
NE	 &	  &0.0033&	&      &      &0.10 &   &   &		&\cr
SW	 &	  &	 &	&      &      &0.11 &   &   &		&\cr
\hline
1209+746 &4CT.74.17.1& G &      &24.99 &23.26 &	    & 1 &   &\cite{zb9864,29}&\cr
NW	&	  &0.107 &      &      &     &120  &   &   &		&\cr
\hline
1216+423 &M106    &  S2  &      &21.68 &19.70 &	    & 2 &   &\cite{c00536675}&\cr
SE	 &NGC4258 &0.002 &      &      &      &3.4  &   &   &		&\cr
NW	 &	  &	 &	&      &      &2.7  &   &   &		&\cr
\hline
1216+061 &3C270   &  G   &      &24.01 &22.25 &	    & 2 & C &\cite{mk931023,jf364213,j00534165}&\cr
E	 &NGC4261 &0.0073&      &      &      &31   &   &   &\cite{19,29}&\cr
W	 &	  &	 &	&      &      &31   &   &   &		&\cr
\hline
1217+023 &PKS     &  Q   &      &25.68 &25.33 &     & 1 &   &\cite{29,74,92}&\cr
NE	 &	  & 0.240&      &      &      &121  &   &   &		&\cr
\hline
1219+285 &        &  BL  &      &25.56 &24.26 &     & 1 &   &\cite{k981151295}&\cr
SE	 &	  & 0.10 &      &      &      &0.011&   &   &		&\cr
\hline
1221+809 &S5	  &  BL	 &	&      &      &	    & 1 &   &\cite{hb1001,211,w98300790}&\cr
N	 &	  &	 &	&      &      &     &   &   &		&\cr
\hline
1222+216 &	  &  Q	 &	&26.64 &26.19 &     & 1 & S &\cite{sw931658}&\cr
NE	 &	  &0.435 &25.93	&      &      &42   &   &   &		&\cr
\hline
1222+13  &3C272.1 &  G   &      &23.24 &21.72 &	    & 2 &   &\cite{29,129}&\cr
N	 &M84     &0.0031&     	&      &      &3.3  &   &   &		&\cr
S	 &        &	 &      &      &      &     &   &   &		&\cr
\hline
1223+129 &NGC4388 &  S1  &      &22.11 &20.16 &	    & 1 & L &\cite{m00529816}&\cr
SW	 &        &0.008 &     	&      &      & 0.35&   &   &		&\cr
\hline
1225+368 &ON343	  &  Q	 &	&28.36 &27.37 &     & 1 &   &\cite{df9527,xr99297,p0053490}&\cr
W	 &	  &1.9747&	&      &      &0.23 &   &   &		&\cr
\hline
1226+023*&3C273   &  Q   &      &27.14 &26.92 &	    & 1 &   &\cite{ac96543,c9051,kb903,bp91l1,rm91458,15}&\cr
SW       &        & 0.158& 26.66&      &      &39   &   &   &\cite{br9231,cg93347,56,rc96414,54}&\cr
	 &	  &	 &	&      &      &	    &   &   &\cite{zu901777,km952305,lz952479,29,35}&\cr
	 &	  &	 &	&      &      &	    &   &   &\cite{du354374,cg96p203,61,69,83}&\cr
	 &	  &	 &	&      &      &	    &   &   &\cite{60,103,104,195,198}&\cr
	 &	  &	 &	&      &      &	    &   &   &\cite{211,49,245,80,125}&\cr
	 &	  &	 &	&      &      &	    &   &   &\cite{97,117,119,86,105}&\cr
	 &	  &	 &	&      &      &	    &   &   &\cite{222,223,55,nm32669,133}&\cr
	 &	  &	 &	&      &      &	    &   &   &\cite{218,219,207,269,270}&\cr
	 &	  &	 &	&      &      &	    &   &   &\cite{t98506637,k981151295,l0054166,m99346397,h991181942}&\cr
	 &	  &	 &	&      &      &	    &   &   &\cite{r0036099}&\cr
\hline
1226+105 &MC2	  &  Q   &  	& 27.81&26.51 &	    & 1 & C &\cite{gc91171,lb8763}&\cr
SW	 &	  & 2.296&27.51	&      &      &8.1  &   &   &\cite{7,8,192}&\cr
\hline
1227+119 &	  &  G	 &	&25.12 &23.34 &     & 2 & L &\cite{ol10841}&\cr
NW	 &	  &0.0843&	&      &      &26   &   &   &		&\cr
SE	 &	  &	 &	&      &      &26   &   &   &		&\cr
\hline
1228+126*&M87     &  S   &      &24.78 &22.92 &     & 1 &   &\cite{j90l8,br9231,bm92393,nm95662,17}&\cr
NW   \%\ &Vir A   &0.0043& 23.27&      &      &2.0  &   &   &\cite{mr9661,df96375,bs911632,87,49}&\cr
	 &3C274	  &	 &	&      &      &	    &   &   &\cite{zm93175,tr95921,of90449,106,82}&\cr
	 &NGC4486 &	 &	&      &      &	    &   &   &\cite{sb473254,sm354132,cm317637,29,52}&\cr
	 &Arp 152 &	 &	&      &      &	    &   &   &\cite{nm318383,sp285181,jb95500,119,188}&\cr
	 &	  &	 &	&      &      &	    &   &   &\cite{bz447582,197,116,128,172}&\cr
	 &	  &	 &	&      &      &	    &   &   &\cite{198,201,202,224,234}&\cr
	 &	  &	 &	&      &      &	    &   &   &\cite{117,122,123,146,193}&\cr
	 &	  &	 &	&      &      &	    &   &   &\cite{222,236,238,254,k981151295}&\cr
	 &	  &	 &	&      &      &	    &   &   &\cite{m9912281,p991172185,j99401891,b00535615,o00543611}&\cr
	 &	  &	 &	&      &      &	    &   &   &\cite{b99520621,s00542667}&\cr
\hline
1231+674 &4C67.21 &  G	 &	& 25.06&22.85 &	    & 2 &   &\cite{167,ol10841,k99516716}&\cr
NE	 &	  &0.1062&23.03	&      &      &16   &   &   &\cite{158,168}&\cr
SW	 &	  & 	 &23.29	&      &      &32   &   &   &		&\cr
\hline
1233+169 &	  &  G	 &	&24.64 &23.60 &     & 2 &   &\cite{ol10841}&\cr
SW	 &	  &0.0784&	&      &      &49   &   &   &		&\cr
SE	 &	  &	 &	&      &      &29   &   &   &		&\cr
\hline
1233+168 &	  &  G   &	&24.96 &22.68 &     & 2 & C &\cite{ol10841}&\cr
NE	 &	  &0.0784&	&      &      &23   &   &   &		&\cr
SE	 &	  &	 &	&      &      &23   &   &   &		&\cr
\hline
1234-723 &B	  &  G   &	&23.87 &23.16 &     & T &   &\cite{l96p227}&\cr
N	 &	  &0.023 &	&      &      &103  &   &   &		&\cr
S(C-J)	 &	  &	 &	&      &      &97   &   &   &		&\cr
\hline
1234+396 &	  &      &	&      &      &     & 1 &   &\cite{k99307225}&\cr
E	 &	  &      &	&      &      &     &   &   &		&\cr
\hline
1235-182 &MC	  &  Q	 &	&27.34 &26.33 &	    & 1 &   &\cite{lb8763}&\cr
E	 &	  &2.192 &	&      &      &13   &   &   &		&\cr
\hline
1238+188 &	  &  G	 &	&24.48 &22.88 &     & 2 & C &\cite{ol10841}&\cr
W	 &	  &0.0718&	&      &      &104  &   &   &		&\cr
E	 &	  &	 &	&      &      &104  &   &   &		&\cr
\hline
1241+166 &3C275.1 &  Q	 &	& 27.08& 25.74&	    & 1 &   &\cite{gc91171,lp92545}&\cr
NW	 &	  & 0.557&26.55	&      &      &36   &   &   &\cite{29,243}&\cr
\hline
1243+267 &	  &  G   &      &24.22 &22.81 &	    & 2 &   &\cite{ow80501}&\cr
N	 & 	  &0.0891& 23.97&      &      &45   &   &   &\cite{175}	&\cr
S	 &	  &	 & 23.12&      &      &50   &   &   &		&\cr
\hline
1244+492 &4C49.25 &  S2	 &	&25.84 &24.30 &     & 1 &   &\cite{ss95629}&\cr
NE	 &	  &0.206 &24.22	&      &      &3.7  &   &   &		&\cr
\hline
1244+699 &	  &  G	 &	&24.80 &23.25 &	    & 2 & C &\cite{162}	&\cr
NE	 &	  &0.157 &23.23	&      &      &26   &   &   &		&\cr
SW	 &	  &	 &23.23	&      &      &26   &   &   &		&\cr
\hline
1247-012 &	  &  G	 &	&24.34 &23.45 &     & 1 & C &\cite{ol10841}&\cr
NW	 &	  &0.0825&	&      &      &34   &   &   &		&\cr
\hline
1250-102 &NGC4760 &  G   &      &23.27 &22.14 &     & 2 &   &\cite{29,210}&\cr
NW	 &        &0.0138& 	&      &      &2.9  &   &   &		&\cr
SE	 &	  &	 &	&      &      &     &   &   &		&\cr
\hline
1250+568 &3C277.1 &  Q   &      &26.61 &25.70 &     & 1 &   &\cite{l98299467}&\cr
NW	 &        & 0.32 & 	&      &      &2.7  &   &   &		&\cr
\hline
1251+273 &NGC4789 &  G   &      &23.55 &21.16 &     & 2 &   &\cite{cf91362}&\cr
NW	 &UGC8028 & 0.027&  	&      &      &6.7  &   &   &\cite{29}	&\cr
SE	 &	  &	 &  	&      &      &6.7  &   &   &		&\cr
\hline
1251+278 &3C277.3 &  G   &      &25.38 &22.99 &     & 1 &   &\cite{29,119,153,282}&\cr
S	 &Com A   &0.0857& 23.52&      &      &11   &   &   &		&\cr
\hline
1252-122 &3C278   &  G   &      &24.23 &22.13 &	    & 2 & L &\cite{mk931023}&\cr
E	 &NGC4783 &0.0138&      &      &      &14   &   &   &\cite{29,21}&\cr
W	 &	  &	 &	&      &      &14   &   &   &		&\cr
\hline
1253-055*&3C279   &  Q   &      &27.56 &27.56 &	    & 1 &   &\cite{br9231,ca9383,al94247,lz952479,29}&\cr
SW	 &	  & 0.536&26.66	&      &      &14   &   &   &\cite{cg96861,35,198,109,192}&\cr
	 &	  &	 &	&      &      &	    &   &   &\cite{244,274,r98131451,p0053791,l0054166}	&\cr
	 &	  &	 &	&      &      &	    &   &   &\cite{l98504702}&\cr
\hline
1254+277 &NGC4839 &  G   &      &22.63 &21.18 &     & 2 &   &\cite{175}&\cr
N	 &B2      &0.0249& 22.2 &      &      &5.4  &   &   &		&\cr
S	 &        &      & 22.06&      &      &6.5  &   &   &		&\cr
\hline
1256+281 &NGC4869 &   G  &      & 22.91& 21.08&	    & 2 & C &\cite{29,162}&\cr
NE	 &5C4.81  &0.0235& 22.02&      &      &3.2  &   &   &		&\cr
SW	 &	  &	 & 22.02&      &      &3.2  &   &   &		&\cr
\hline
1257+28  &NGC4874 &   G	 &	& 23.05& 20.81&	    & 2 &   &\cite{162,77}&\cr
NE	 &5C4.85  &0.0232& 	&      &      &3.2  &   &   &		&\cr
SW	 &	  &	 &     	&      &      &3.2  &   &   &		&\cr
\hline
1258-321 &PKS     &   G	 &	& 23.53&22.54 &	    & 2 &   &\cite{dw28485}&\cr
NW	 &B	  &0.017 &	&      &      &6.6  &   &   &\cite{29}	&\cr
SE	 &	  &	 &	&      &      &6.6  &   &   &		&\cr
\hline
1258+404 &3C280.1 &   Q	 &	& 27.84& 26.21&	    & 1 & B &\cite{al94247,gc91171,lb8763}&\cr
SE	 &	  & 1.659& 27.65&      &      &98   &   &   &\cite{29,240}&\cr
NW(C-J)	 &	  & 	 &	&      &      &     &   &   &		&\cr
\hline
1303-250 &MRC     &   Q	 &	& 26.63&25.46 &	    & 1 & C &\cite{k98118327}&\cr
NW	 &B2	  & 0.738&      &      &      &73   &   &   &		&\cr
\hline
1305+801 &S5	  &  Q	 &	&27.15 &26.41 &	    & 1 &   &\cite{w98300790}&\cr
NE	 &        &1.183 &	&      &      &75   &   &   &		&\cr
\hline
1306+107 &4C10.35 &   G	 &	& 24.33&23.12 &	    & 2 & C &\cite{ow80501}&\cr
NW	 &	  & 0.136&22.96 &      &      &13   &   &   &\cite{167}	&\cr
SE	 &	  & 	 &	&      &      &13   &   &   &		&\cr
\hline
1308+182 &4C18.36 &   Q	 &	& 27.32& 26.00&	    & 1 & C &\cite{gc91171,lb8763}&\cr
SE	 &	  & 1.689&26.89	&      &      &16   &   &   &\cite{7,239}&\cr
NW(C-J)	 &	  & 	 &	&      &      &     &   &   &		&\cr
\hline
1308-441 &	  &   G	 &	&24.55 &23.81 &     & ? &   &\cite{jm80137}&\cr
NW	 &	  &0.051 &	&      &      &245  &   &   &		&\cr
\hline
1311-270 &PKS	  &   Q	 &	& 27.8 &26.66 &	    & 1 &   &\cite{lb8763}&\cr
SW	 &	  & 2.195&	&      &      &42   &   &   &\cite{7}	&\cr
NE(C-J)  &	  &	 &	&      &      &38   &   &   &		&\cr
\hline
1313+073 &	  &   G	 &	& 24.75& 22.57&     & 2 & L &\cite{177,178}&\cr
E	 &	  &0.0507&	&      &      &25   &   &   &		&\cr
W	 &	  &	 &	&      &      &--   &   &   &		&\cr
\hline
1315+346 &B2      &   Q  &      & 26.98& 26.76&	    & 1 & C &\cite{29,w98300790}&\cr
NE	 &OP326   & 1.050&  	&      &      &13   &   &   &		&\cr
\hline
1316+299 &4C29.47 &   G  &      & 24.85& 23.25&     & 2 & S &\cite{29,57,175}&\cr
E	 &        &0.0728&      &      &      &  110&   &   &		&\cr
W	 &	  &	 &	&      &      &  110&   &   &		&\cr
\hline
1317+520 &4C52.27 &   Q  &      & 27.37& 26.78&	    & 1 & C &\cite{gc91171,29,174,170}&\cr
SE	 &	  & 1.060&26.91	&      &      & 41  &   &   &		&\cr
\hline
1318-434 &NGC5090 &   G  &	&23.87 &22.88 &     & 2 & S &\cite{mk931023,lj961197}&\cr
NE	 &PKS	  &0.011 &	&      &      &29   &   &   &\cite{l96p227}&\cr
SW	 &B	  &	 &	&      &      &24   &   &   &		&\cr
\hline
1318+113 &4C11.45 &   Q	 &	& 28.35& 26.88&	    & 1 & C &\cite{gc91171,lb8763}&\cr
SW	 &	  & 2.171& 27.88&      &      &44   &   &   &\cite{7,143}&\cr
\hline
1319+428 &3C285   &   G	 &	& 25.27&22.69 &	    & 1 &   &\cite{132}	&\cr
NE	 &	  &0.0794&23.76	&      &      &84   &   &   &		&\cr
\hline
1320+584 &A1731	  &  G	 &	& 25.15&23.56 &     & 1 &   &\cite{ol10841}&\cr
E	 &	  &0.1932&	&      &      &23   &   &   &		&\cr
N(C-J)	 &	  &	 &	&      &      &23   &   &   &		&\cr
\hline
1321+31  &NGC5127 &   G  &      & 23.85& 21.77&	    & 2 & S &\cite{29,67}&\cr
SE	 &B2	  &0.0161& 	&      &      &55   &   &   &		&\cr
NW	 &	  &	 &	&      &      &	    &   &   &		&\cr
\hline
1322-427\%&Cen A  &   G  &      & 24.62& 22.12&     & 1 &   &\cite{jh9182,cb395444,dj470l15,26,29}&\cr
NE	 &NGC5128 &0.0012& 21.97&      &      &5.2  &   &   &\cite{jt96p23,a94738,30,32,62}&\cr
SW(C-J)  &PKS	  &	 &20.32	&      &      &	    &   &   &\cite{119,45,75,155,pt96p21}&\cr
	 &	  &	 &	&      &      &	    &   &   &\cite{169,t98115960,m99307750,f00521021,k005319}&\cr
\hline
1322+36  &NGC5141 &   G  &      & 23.42& 22.38&	    & T &   &\cite{175}	&\cr
S	 &4C36.27 &0.0175& 22.45&      &      &7    &   &   &		&\cr
N(C-J)	 &B2	  &	 &	&      &      &7    &   &   &		&\cr
\hline
1323+655 &4C65.15 &   Q  &      & 27.44& 25.71&	    & 1 & L &\cite{gc91171,lb8763}&\cr
SW	 &	  & 1.618& 	&      &      &29   &   &   &\cite{7,228}&\cr
\hline
1323+321 &	  &   G  &      & 27.22&  --  &	    & 1 &   &\cite{k981151295}&\cr
SE	 &	  & 0.37 & 	&      &      &0.078&   &   &		&\cr
\hline
1328+307 &3C286   &   Q  &      & 28.18& 27.88&	    & 1 & B &\cite{al94247,kw901057,ss91225,29}&\cr
SW	 &        & 0.849& 26.68&      &      &0.54 &   &   &\cite{zs9432,143,232}&\cr
E(C-J)	 &	  &	 & 26.49&      &      &0.17 &   &   &		&\cr
\hline
1330+022 &3C287.1 &   Q  &      & 26.19& 25.26&	    & 1 &   &\cite{h9811925}&\cr
W	 &        &0.2159& 	&      &      &61   &   &   &		&\cr
\hline
1331-099 &        &   G  &      & 25.23& 23.61&	    & 1 &   &\cite{s9732878}&\cr
SE	 &        & 0.081& 22.71&      &      &0.63 &   &   &		&\cr
\hline
1333-337 &IC4296  &   G  &      & 24.41& 22.46&     & 2 & C &\cite{mk931023,jm80137}&\cr
NW	 &	  &0.0129&23.29 &      &      &81   &   &   &\cite{19,29,68,120,121}&\cr
SE	 &        &      &23.29 &      &      &81   &   &   &		&\cr
\hline
1335+047 &NGC5252 &  S2	 &	&21.45 &21.68 &     & ? & C &\cite{m00529816}&\cr
NW	 &	  &0.022 &	&      &      &9.5  &   &   &		&\cr
\hline
1335+552 &	  &  Q	 &	&27.11 &26.96 &     & 1 &   &\cite{tv95345}&\cr
SE	 &	  &1.096 &	&      &      &0.059&   &   &		&\cr
\hline
1336+39  &3C288   &   G	 &	& 26.39& 23.85&	    & T &   &\cite{27}	&\cr
NW	 &        & 0.246& 24.63&      &      &6    &   &   &		&\cr
SE	 &	  &	 & 23.68&      &      &-    &   &   &		&\cr
\hline
1343-601 &Cen B	  &  G	 &	&25.20 &23.58 &     & 2 &   &\cite{m91255,j96p221}&\cr
SW	 &	  &0.01215&	&      &      &66   &   &   &		&\cr
NE(C-J)	 &	  &	 &	&      &      &45   &   &   &		&\cr
\hline
1345+125 &4C12.50 &  S2	 &	&26.10 &24.87 &     & 1 &   &\cite{fc105299,sod325943}&\cr
SE	 &PKS	  &0.121 &	&      &      &0.11 &   &   &		&\cr
\hline
1345+584 &4C58.27 &  Q	 &	& 27.61&26.41 &     & 1 & C &\cite{lb8763}&\cr
W	 &	  &2.039 &	&      &      &52   &   &   &		&\cr
\hline
1347+539 &4C53.28 &  Q	 &	&27.16 &26.77 &     & 1 &   &\cite{pw981,xr99297,fc11195}&\cr
SE	 &	  &0.978 &	&      &      &0.12 &   &   &		&\cr
\hline
1350+316 &3C293   &   G	 &	&25.01 &22.70 &     & T &   &\cite{al961,e99511730}&\cr
E	 &	  &0.0452&	&      &      &1.2  &   &   &		&\cr
W	 &	  &	 &	&      &      &0.36 &   &   &		&\cr
\hline
1354+195 &PKS     &   Q	 &	& 27.15& 26.92&	    & 1 &   &\cite{gc91171,mb93298,s98298877}&\cr
SE	 &4C19.44 & 0.720&26.45	&      &      &66   &   &   &\cite{209}	&\cr
\hline
1354+258 &PKS	  &  Q	 &	& 27.32&26.44 &     & 1 & L &\cite{lb8763}&\cr
NE	 &OP291	  &2.032 &	&      &      &41   &   &   &		&\cr
\hline
1355+219 &	  &  G	 &	&23.74 &23.47 &     & 2 &   &\cite{ol10841}&\cr
NW	 &	  &0.0668&	&      &      &38   &   &   &		&\cr
SE	 &	  &	 &	&      &      &38   &   &   &		&\cr
\hline
1357+28  &        &   G  &      & 24.03& 22.45&	    & 2 &   &\cite{175}	&\cr
N	 & 	  &0.0629& 23.44&      &      &23   &   &   &		&\cr
S	 &	  &	 & 23.00&      &      &22   &   &   &		&\cr
\hline
1358+624 &4C62.22 &  S	 &	& 27.01&24.22 &     & 1 &   &\cite{df9527,tr46395}&\cr
SE	 &	  &0.431 &	&      &      &0.17 &   &   &		&\cr
\hline
1359-113 &	  &  G	 &	&24.49 &23.08 &	    & 1 &   &\cite{ow80501}&\cr
NW	 &	  &0.0365&	&      &      &97   &   &   &		&\cr
\hline
1400-001 &	  &  G	 & 	&27.29 &25.51 &     & 1 &   &\cite{cr1091}&\cr
NW	 &	  &2.363 & 	&      &      &34   &   &   &		&\cr
\hline
1400+162 &MC	  &   BL & 	& 25.76& 25.01&	    & 1 &   &\cite{92,215}&\cr
NE	 &	  & 0.244& 	&      &      &6    &   &   &		&\cr
\hline
1402+044 &PKS	  &  Q	 &	&28.46 &28.22 &     & 1 & L &\cite{fc11195}&\cr
NW	 &	  &3.211 &	&      &      &0.063&   &   &		&\cr
\hline
1407+177 &NGC5490 &   G  &      & 23.68& 22.00&	    & 2 &   &\cite{cf91362}&\cr
E	 &UGC9058 &0.0163& 	&      &      &2.0  &   &   &\cite{29}	&\cr
W	 &	  &	 & 	&      &      &5.5  &   &   &		&\cr
\hline
1409+595 &	  &   Q	 &	&      &      &	    & 1 &   &\cite{pb92655}&\cr
E	 &	  &	 &	&      &      &	    &   &   &		&\cr
\hline
1413+135 &PKS	  &  BL	 &	&25.91 & 25.61&	    & 1 &   &\cite{pc961839,fc11195,p0053490}&\cr
SW	 &	  &0.249 &	&      &      &0.23 &   &   &		&\cr
NE(C-J)	 &	  &	 &	&      &      &0.084&   &   &		&\cr
\hline
1413+349 &	  &	 &	&      &      &	    & 1 &   &\cite{df9527}&\cr
NE	 &	  &	 &	&      &      &	    &   &   &		&\cr
SW(C-J)	 &	  &	 &	&      &      &	    &   &   &		&\cr
\hline
1414+110 &3C296   &   G  &      & 24.43& 22.69&	    & 2 &   &\cite{lp91537,ha288l1}&\cr
SW	 &NGC5532 &0.0237&      &      &      &50   &   &   &\cite{20,29}&\cr
NE	 &        &	 &	&      &      &50   &   &   &		&\cr
\hline
1415+253 &NGC5548 &  S1.5&	&21.86 &21.07 &     & 2 & S &\cite{n99120209,w00531716}&\cr
NW	 &UGC9149 &0.017 &	&      &      &1.2  &   &   &		&\cr
SE	 &	  &	 &	&      &      &0.23 &   &   &		&\cr
\hline
1415+084 &	  &  G	 &	&24.08 &22.28 &     & 2 &   &\cite{ol10841}&\cr
NE	 &	  &0.0570&	&      &      &57   &   &   &		&\cr
SW	 &	  &	 &	&      &      &14   &   &   &		&\cr
\hline
1416+423 &3C298	  &  Q	 &	&28.27 &27.01 &	    & 1 &   &\cite{g84p43,l98299467}&\cr
E	 &	  &1.44	 &	&      &      &4.6  &   &   &		&\cr
\hline
1418+546 &OQ530   &  BL  &	& 25.33&25.39 &     & 1 &   &\cite{pw981}&\cr
SE	 &	  &0.152 &	&      &      &0.072&   &   &		&\cr
\hline
1420+198 &3C300	  &  G	 &	& 26.44&23.74 &     & 1 &   &\cite{ha288859}&\cr
NW	 &	  &0.272 &25.03	&      &      &182  &   &   &		&\cr
\hline
1422+202 &4C20.33 &   Q	 &	& 27.29& 25.56&	    & 1 &   &\cite{mj92353,m97125573,b97125453}&\cr
S	 & 	  & 0.871&26.94	&      &      &24   &   &   &\cite{74}	&\cr
\hline
1422+26  &B2	  &   G  &	& 24.00& 22.25&	    & 2 &   &\cite{175,284,285}&\cr
SE	 &	  &0.0370& 22.25&      &      &13   &   &   &		&\cr
NW	 &	  &	 & 22.35&      &      &15   &   &   &		&\cr
\hline
1433+177 &4C17.59 &   Q	 &	& 27.16&      &	    & 1 &   &\cite{211}	&\cr
SE	 &	  & 1.203&	&      &      &22   &   &   &		&\cr
\hline
1433+553 &4C55.29 &   G	 &	& 25.05&23.44 &     & 1 &   &\cite{ol10841,k99516716}&\cr
NW(C-J)	 &	  &0.1396&22.85	&      &      &45   &   &   &\cite{167}	&\cr
SE	 &	  &	 &23.36	&      &      &45   &   &   &		&\cr
\hline
1433+304 &	  &	 &	&      &      &     & 1 &   &\cite{fc11195}&\cr
NE	 &	  &	 &	&      &      &     &   &   &		&\cr
\hline
1441+522 &3C303   &   S1 &      & 25.75& 24.53&	    & 1 &   &\cite{al94247,lp91537}&\cr
NW	 &	  & 0.141& 24.33&      &      &26   &   &   &\cite{29,129,126,144}&\cr
\hline
1447+771 &3C305.1 &   G  &      & 27.06& 27.09&	    & 1 &   &\cite{l98299467}&\cr
NE	 &	  & 1.13 & 	&      &      &1.1  &   &   &		&\cr
\hline
1448+634 &3C305   &   G  &      & 24.73& 22.57&     & 2 & S &\cite{js95339,lb96p157}&\cr
NE	 &	  & 0.041& 	&      &      &1.4  &   &   &\cite{29}	&\cr
SW(C-J)  &	  &	 & 	&      &      &1.1  &   &   &		&\cr
\hline
1450+28  &B2      &   G  &      & 24.56& 22.56&	    & 2 &   &\cite{29,175}&\cr
E	 &	  &0.1265& 23.81&      &      &15   &   &   &		&\cr
W	 &	  &	 & 21.89&      &      &25   &   &   &		&\cr
\hline
1450+333 &B       &   G	 &	&      & 23.46&     & 1 &   &\cite{s00315371}&\cr
N	 &        & 0.249&      &      &      &51   &   &   &		&\cr
S(C-J)	 &        &      &      &      &      &52   &   &   &		&\cr
\hline
1451-375 &PKS     &   Q	 &	& 26.36& 26.24&     & 1 &   &\cite{29,s98298877}&\cr
SW	 &        & 0.314&      &      &      &17   &   &   &		&\cr
NE(C-J)	 &        &      &      &      &      &     &   &   &		&\cr
\hline
1452-517 &	  &   G	 &	&23.75 &22.68 &	    & 2 &   &\cite{jm80137,l96p227}&\cr
NE	 &	  &0.016 &	&      &      &64   &   &   &		&\cr
SW	 &	  &	 &	&      &      &64   &   &   &		&\cr
\hline
1453+120 &UGC9602 &   G	 &	&23.65 &22.77 &     & 2 &   &\cite{cf91362}&\cr
NE	 &PGC53379&0.0323&	&      &      &59   &   &   &		&\cr
SW	 &	  &	 &	&      &      &43   &   &   &		&\cr
\hline
1458+718 &3C309.1 &   Q  &      & 28.01& 27.62&     & 1 & C &\cite{29,198,255,fc11195,l98299467}&\cr
NE	 &4C71.15 & 0.904& 26.60&      &      &3.8  &   &   &		&\cr
\hline
1508+065 &	  &   G	 &	&24.61 &22.98 &     & 2 &   &\cite{ol10841}&\cr
SE	 &	  &0.0817&	&      &      &10   &   &   &		&\cr
\hline
1508-055 &	  &   Q	 &	&28.32 &26.88 &     & 1 &   &\cite{k981151295}&\cr
E	 &	  & 1.18 &	&      &      &0.087&   &   &		&\cr
\hline
1508+059 &PKS	  &   G  &	& 25.13&22.52 &     & 2 &   &\cite{tb941942}&\cr
NW	 &	  &0.0767&	&      &      &10   &   &   &		&\cr
SE	 &	  &	 &	&      &      &10   &   &   &		&\cr
\hline
1509+158 &4C15.45 &  Q	 &	&26.96 &26.08 &	    & 1 &   &\cite{sj90408,pg86365}&\cr
SW	 &	  &0.828 &25.80	&      &      &33   &   &   &\cite{212}	&\cr
\hline
1510-089A&PKS	  &  Q	 &	&28.76 &28.19 &     & 1 &   &\cite{fc11195}&\cr
NW	 &	  &2.10	 &	&      &      &0.14 &   &   &		&\cr
\hline
1510-089 &PKS     &   Q	 &	& 26.41& 26.40&	    & 1 &   &\cite{160}	&\cr
SE	 &	  & 0.361&  	&      &      &24   &   &   &		&\cr
\hline
1511+103 &MC2	  &   Q  &	&26.64 &25.49 &     & 1 &   &\cite{lb8763}&\cr
NE	 &	  &1.546 &	&      &      &95   &   &   &		&\cr
\hline
1512+370 &PG	  &   Q	 &	&26.18 &      &	    & 1 &   &\cite{253} &\cr
	 &	  & 0.371&	&      &      &     &   &   &		&\cr
\hline
1514-241 &Ap Lib  &  BL  &	&24.79 &24.68 &     & 1 & C &\cite{sd325911,l98504702}&\cr
NE	 &	  &0.049 &	&      &      &45   &   &   &		&\cr
\hline
1518+045 &PKS	  &   G  &	&24.91 &23.85 &     & 2 &   &\cite{dw28485}&\cr
NE	 &	  &0.052 &	&      &      &47   &   &   &		&\cr
SW	 &	  &	 &	&      &      &47   &   &   &		&\cr
\hline
1519+567 &	  &	 &	&      &      &	    & 2 &   &\cite{pb92655}&\cr
E	 &	  &	 &	&      &      &	    &   &   &		&\cr
W(C-J)	 &	  &	 &	&      &      &	    &   &   &		&\cr
\hline
1521+288 &4C28.39 &   G  &      & 24.58& 23.58&	    & 1 &   &\cite{cf95643}&\cr
SE	 &B2	  &0.0825& 23.80&      &      &53   &   &   &\cite{175}	&\cr
NW(C-J)	 &	  &	 &	&      &      &     &   &   &		&\cr
\hline
1522+155 &MC3	  &   Q	 &	&26.36 &26.24 &	    & 1 &   &\cite{au84617}&\cr
S	 &	  &0.628 &	&      &      &39   &   &   &		&\cr
\hline
1525+290 &        &   G	 &	& 23.98& 22.10&	    & T &   &\cite{bb9353,175,bb9353}&\cr
NE	 &	  &0.0653& 23.29&      &      &11   &   &   &		&\cr
SW	 &	  &	 & 23.40&      &      &11   &   &   &		&\cr
\hline
1528+29  &	  &   G  &  	& 24.21& 22.31&     & 2 &   &\cite{175}	&\cr
NE	 &	  &0.0843& 23.48&      &      &130  &   &   &		&\cr
SW	 &        &      & 23.24&      &      &88   &   &   &		&\cr
\hline
1529+242 &3C321   &  S2	 &	& 26.15&23.62 &	    & 1 &   &\cite{21,132}&\cr
NW	 &	  &0.0960&	&      &      &16   &   &   &		&\cr
SE(C-J)	 &	  &      &	&      &      &     &   &   &		&\cr 
\hline
1532+016 &	  &  Q	 &	& 27.78&27.07 &	    & 1 &   &\cite{k981151295}&\cr
SE	 &	  & 1.44 &	&      &      &0.024&   &   &		&\cr
\hline
1540+180 &4C18.43 &   Q	 &	& 27.93&26.91 &	    & 1 &   &\cite{lb8763,7}&\cr
N	 &	  &1.662 &	&      &      &15   &   &   &		&\cr
\hline
1545+210 &3C323.1 &  Q	 &	&26.28 &24.50 &     & 1 &   &\cite{bh9491}&\cr
SW	 &	  &0.264 &	&      &      &38   &   &   &		&\cr
\hline
1547+215 &3C324	  &  Q	 &	&27.87 &23.58 &     & 1 &   &\cite{b98299357}&\cr
E	 &	  &1.206 &	&      &      &30   &   &   &		&\cr
\hline
1549+628 &3C325	  &  G	 &	&      &25.23 &     & 2 &   &\cite{g00544659}&\cr
NW	 &	  &0.86  &	&      &      &78   &   &   &		&\cr
SE(C-J)	 &	  &      &	&      &      &98   &   &   &		&\cr
\hline
1550+703 &	  &  	 &	&      &      &     & 1 & S &\cite{k99307225}&\cr
SE	 &	  &      &	&      &      &     &   &   &		&\cr
\hline
1553+24  &	  &   G  &	& 23.36& 23.01&	    & 2 &   &\cite{154,175,176,c00362871}&\cr
NW	 &	  &0.0426& 22.93&      &      &21   &   &   &		&\cr
SE	 &	  &      & 22.69&      &      &17   &   &   &		&\cr
\hline
1555+332 &GC 	  &   Q	 &	& 26.38& 23.98&	    & 1 &   &\cite{74}	&\cr
W	 &	  & 0.942& 	&      &      &80   &   &   &		&\cr
\hline
1559+021 &3C327	  &  S2	 &	&26.10 &23.39 &     & 1 &   &\cite{lb29120}&\cr
W	 &	  &0.1039&22.22	&      &      &228  &   &   &		&\cr
E(C-J)	 &	  &	 &	&      &      &     &   &   &		&\cr
\hline
1600+243 &	  &  	 &	&      &      &	    & 1 &   &\cite{w98300790}&\cr
NE	 &        &      &	&      &      &     &   &   &		&\cr
\hline
1602-001 &4C-00.63&   Q	 &	& 27.66&27.01 &	    & 1 &   &\cite{7,lb8763}&\cr
W	 &  	  & 1.625&	&      &      &68   &   &   &		&\cr
\hline
1603+178 &NGC6047 &   G  &	&23.93 &21.97 &     & 2 & S &\cite{go941523,fg8821}&\cr
N	 &	  &0.0319&22.68	&      &      &6.1  &   &   &		&\cr
S(C-J)	 &	  &	 &	&      &      &2.6  &   &   &		&\cr
\hline
1603+001 &4C00.58 &   G	 &	& 24.91&23.27 &     & 1 & S &\cite{b99310223}&\cr
E	 &	  &0.059 &	&      &      &9.8  &   &   &		&\cr
\hline
1606+180 &4C18.47 &   Q	 &	& 26.28&24.34 &     & 1 &   &\cite{pg86365}&\cr
SW	 &	  &0.346 &	&      &      &27   &   &   &		&\cr
\hline
1607+268 &CTD93	  &   G	 &	&26.99 &25.78 &     & 1 &   &\cite{fc11195,s99515558}&\cr
SW	 &PKS	  &0.473 &	&      &      &0.18 &   &   &		&\cr
\hline
1610-608 &B	  &   G	 &	&23.97 & 22.41&     & 2 & L &\cite{jm96137,j96p221}&\cr
E	 &	  &0.0143&	&      &      &49   &   &   &		&\cr
W	 &	  &	 &	&      &      &49   &   &   &		&\cr
\hline
1611+343 &DA 406  &   Q	 &	&27.88 & 27.78&     & 1 & L &\cite{l98504702,b991221,l0054166}&\cr
S	 &	  &1.401 &	&      &      &0.020&   &   &		&\cr
\hline
1613+27  &B2	  &   G	 &	& 24.03& 22.69&	    & 2 &   &\cite{175}	&\cr
S	 &	  &0.0647& 23.24&      &      &13   &   &   &		&\cr
N	 &	  &	 & 23.04&      &      &11   &   &   &		&\cr
\hline
1615+425 &	  &   G	 &	& 24.20& 23.20&	    & 2 &   &\cite{29,162}&\cr
NW	 &	  & 0.131&  	&      &      &26   &   &   &		&\cr
SE	 &	  &	 &	&      &      &17   &   &   &		&\cr
\hline
1615+351 &NGC6109 &   G	 &	& 23.40& 22.48&	    & 2 & C &\cite{67,162,175}&\cr
NW	 &B2	  &0.0295&22.82 &      &      &4.3  &   &   &		&\cr
SE(C-J)	 &	  &	 &22.82 &      &      &4.3  &   &   &		&\cr
\hline
1616+366 &	  &	 &	&      &      &	    & 1 &   &\cite{pb92655}&\cr
SW	 &	  &	 &	&      &      &	    &   &   &		&\cr
\hline
1618+177 &3C334   &   Q	 &	& 26.97& 25.78&     & 1 & L &\cite{bh94766,gc91171,107}&\cr
SE	 &	  & 0.555& 25.68&      &      &63   &   &   &\cite{23,24,29,dtb289753}&\cr
NW(C-J)	 &	  &	 & 24.48&      &      &96   &   &   &\cite{bh9491}&\cr
\hline
1621+380 &NGC6137 &   G	 &	& 23.62& 22.75&	    & 2 &   &\cite{162,176}&\cr
S	 &	  &0.0310&	&      &      &1.9  &   &   &		&\cr
N	 &	  &	 &	&      &      &1.9  &   &   &		&\cr
\hline
1622+238 &3C336   &   Q  &	& 27.31& 25.56&	    & 1 &   &\cite{bh94766,24,dtb289753}&\cr
SW	 &	  & 0.927& 26.27&      &      &29   &   &   &		&\cr
C-J	 &	  &	 & 24.97&      &      &65   &   &   &		&\cr
\hline
1622-297 &PKS	  &   Q	 &	& 27.26& 27.18&	    & 1 &   &\cite{t98500673}	&\cr
W	 &	  &0.815 &      &      &      &0.12 &   &   &		&\cr
\hline
1623+410 &	  &   G	 &	& 23.11& 22.76&	    & 1 &   &\cite{267}	&\cr
SE	 &	  &0.0304& 22.74&      &      &0.63 &   &   &		&\cr
\hline
1624+416 &4C41.32 &   Q	 &	&28.57 &27.99 &     & 1 &   &\cite{fc11195}&\cr
N	 &	  &2.55  &	&      &      &0.23 &   &   &		&\cr
\hline
1626+396 &NGC6166 &   G	 &	& 24.51& 23.16&	    & 2 & C &\cite{go941523,fc408446,g98493632,o9849373}&\cr
E	 &3C338   &0.0303&	&      &      &10   &   &   &\cite{175,176,285}&\cr
W	 &B2	  &	 &	&      &      &11   &   &   &		&\cr
\hline
1626+278 &3C341   &   G	 &	& 26.80& 23.49&	    & T &   &\cite{lp91537,29}&\cr
NE	 &	  & 0.448&  	&      &      &113  &   &   &		&\cr
SW	 &	  &	 &	&      &      &142  &   &   &		&\cr
\hline
1629+680 &4C68.18 &  Q	 &	& 28.17& 27.69&	    & 1 & L &\cite{d98129219}&\cr
S	 &	  & 2.475&27.74	&      &      &0.57 &   &   &		&\cr
\hline
1633+382*&4C38.41 &   Q	 &      &28.19 & 28.01&	    & 1 &   &\cite{pw981,bc444l21,l0054166}&\cr
W	 &B2	  & 1.814&  	&      &      &0.37 &   &   &\cite{156,160}&\cr
\hline
1636+379 &4C37.48 &   G  &	& 25.26&23.12 &	    & 1 &   &\cite{167,ol10841}&\cr
NE	 &B2	  & 0.161&23.24	&      &      &62   &   &   &\cite{138}	&\cr
SW(C-J)	 &	  &	 &	&      &      &62   &   &   &		&\cr
\hline
1637+299 &B2	  &   G	 &	& 24.40&22.89 &	    & 2 & C &\cite{283,d98337711}&\cr
SE	 &	  &0.0875&	&      &      &158  &   &   &		&\cr
NW	 &	  &	 &	&      &      &105  &   &   &		&\cr
\hline
1637+826 &NGC6251 &  S2  &	&24.14 &23.66 &	    & 1 &   &\cite{km94729,jw427221,mk123423,29,s0052989}&\cr
NW	 &	  &0.0230&24.11	&      &      &161  &   &   &\cite{114,115,198,112,119}&\cr
SE(C-J)	 &	  &	 &22.46	&      &      &     &   &   &\cite{189,mk324870,n99120209,s00120697}&\cr
\hline
1638+32  &B2	  &   G	 &	& 24.80& 24.06&	    & 1 &   &\cite{175}	&\cr
W	 &	  &0.1398& 24.06&      &      &27   &   &   &		&\cr
\hline
1638+398 &NRAO 512&   Q	 &	&27.66 &27.54 &     & 1 &   &\cite{kw901057}&\cr
SE	 &	  &1.666 &25.81	&      &      &16   &   &   &		&\cr
NW(C-J)	 &	  &	 &	&      &      &     &   &   &		&\cr
\hline
1638+538 &4C53.37 &   G	 &	& 24.93& 23.21&	    & 2 & C &\cite{ol10841,29,48}&\cr
NW	 &	  &0.1098&22.23 &      &      &43   &   &   &		&\cr
SE	 &        &      &22.23	&      &      &27   &   &   &		&\cr
\hline
1641+399*&3C345   &   Q	 &	& 27.57& 27.57&	    & 1 &   &\cite{c9051,br9231,kw93375,rb9544,29}&\cr
NW	 &	  & 0.594& 26.37&      &      &9.5  &   &   &\cite{mp96738,lz952479,br437108,16,37}&\cr
SE(C-J)	 &	  &	 & 24.90&      &      &---  &   &   &\cite{uw432103,zc44335,35,54,55}&\cr
	 &	  &	 &	&      &      &     &   &   &\cite{109,143,181,192,195}&\cr
	 &	  &	 &	&      &      &     &   &   &\cite{198,241,246,124,t98506637}&\cr
	 &	  &	 &	&      &      &     &   &   &\cite{l99521509,g99303515,r0035455,k981151295}&\cr
\hline
1641+173 &3C346	  &  S2	 &	&26.05 &24.81 &	    & 1 &   &\cite{ss91225,dkb107621,k981151295}&\cr
NE	 &	  &0.16	 &25.42	&      &      &6.4  &   &   &		&\cr
\hline
1642+690*&4C69.21 &   Q  & 	& 27.07& 27.08&	    & 1 & C &\cite{kw901057,29,109,160,161}&\cr
S	 & 	  & 0.751&26.25	&      &      &16   &   &   &\cite{180,183,192,vp325484}&\cr
	 &	  &	 &	&      &      &     &   &   &\cite{fc11195,vp325484,k981151295,s98298877}&\cr
\hline
1643+113 &B2	  &      & 	&      &      &	    & 1 & C &\cite{b98293257}&\cr
SE	 &	  &	 &      &      &      &     &   &   &		&\cr
\hline
1643+27  &B2	  &   G  & 	& 24.05& 22.74&	    & 1 &   &\cite{175}	&\cr
N	 &	  &0.1017& 23.15&      &      &18   &   &   &		&\cr
\hline
1648+050 &3C348   &   G	 &	& 27.10& 23.61&	    & 1 & C &\cite{29,99,h9811925,s9935025}&\cr
E	 &Her A   & 0.154&  	&      &      &118  &   &   &		&\cr
W(C-J)	 &	  &	 &	&      &      &112  &   &   &		&\cr
\hline
1652+398 &MRK501  &   BL &	& 24.30& 23.69&     & 1 & B &\cite{pw981,cw43998,216,k981151295,e00521015}&\cr
SE	 &4C39.49 &0.0337&	&      &      &0.016&   &   &\cite{fc11195}&\cr
	 &DA426	  &	 &	&      &      &     &   &   &		&\cr
\hline
1656+571 &	  &  Q	 &	&27.32 &26.99 &     & 1 &   &\cite{tv95345}&\cr
NE	 &	  &1.290 &	&      &      &0.077&   &   &		&\cr
\hline
1658+575 &4C57.29 &   Q	 &	& 27.79&26.48 &	    & 1 &   &\cite{lb8763,7}&\cr
NE	 &	  & 2.173&	&      &      &13   &   &   &		&\cr
\hline
1658+326 &4C32.52E&   G	 &	& 24.52& 22.62&	    & 1 &   &\cite{162}	&\cr
S	 &        &0.1024& 	&      &      &16   &   &   &		&\cr
\hline
1658+30A &B2	  &   G	 &	& 23.88& 22.89&	    & 1 &   &\cite{175,c00362871}&\cr
SW	 &4C30.31 &0.0351& 23.17&      &      &24   &   &   &		&\cr
\hline
1702+298 &4C29.50 &   Q	 &	& 27.93& 26.41&     & 1 &   &\cite{lb8763,7,142}&\cr
NE	 &	  & 1.927& 26.80&      &      &4.2  &   &   &		&\cr
\hline
1704+608 &3C351	  &   Q	 &	& 26.65& 24.14&	    & 1 &   &\cite{ks941163,bh94766}&\cr
NE	 & 	  & 0.371& 24.74&      &      &75   &   &   &\cite{24,dtb289753}&\cr
SW(C-J)	 &	  &	 & 22.85&      &      &2.2  &   &   &		&\cr
\hline
1705+456 &4C45.34 &   Q	 &	&26.84 &26.25 &     & 1 &   &\cite{fc11195}&\cr
SW	 &	  &0.648 &	&      &      &0.050&   &   &		&\cr
\hline
1705+786 &	  &   G  &	&23.32 &21.84 &     & 2 & C &\cite{162}	&\cr
E	 &	  &0.0561&22.72	&      &      &6.2  &   &   &		&\cr
W	 &	  &	 &22.72	&      &      &6.2  &   &   &		&\cr
\hline
1709+460*&3C352	  &   G	 &	&27.23 &24.39 &	    & 1 &   &\cite{gc91171}&\cr
NW	 &	  &0.806 &27.00	&      &      &42   &   &   &		&\cr
\hline
1712+638 &	  &   G  &	& 24.27& 24.33&	    & 2 & C &\cite{162}	&\cr
W	 &	  &0.0829&	&      &      &43   &   &   &		&\cr
E	 &	  &	 &	&      &      &43   &   &   &		&\cr
\hline
1717-009 &3C353	  &   G	 &	&25.70 &      &	    & 2 &   &\cite{sb96p299,s9850729}&\cr
NE	 &	  &0.0304&23.46	&      &      &40   &   &   &		&\cr
SW(C-J)  &	  &	 &22.99	&      &      &43   &   &   &		&\cr
\hline
1719+357 &OT332	  &  Q	 &	&25.60 &25.25 &     & 1 &   &\cite{xr99297}&\cr
S	 &	  &0.263 &	&      &      &183  &   &   &		&\cr
\hline
1721+343 &4C34.47 &   Q	 &	& 25.84& 25.22&	    & 1 &   &\cite{mo912026,9}&\cr
SE	 &B2	  & 0.206& 	&      &      &200  &   &   &		&\cr
\hline
1732+160*&4C16.49 &   Q	 &	&27.52 &25.40 &	    & 1 &   &\cite{gc91171,lb8763}&\cr
N	 &	  &1.270 &27.04 &      &      &63   &   &   &		&\cr
\hline
1736+32  &B2	  &   G	 &	& 24.14& 22.92&	    & 1 &   &\cite{175} &\cr
NW	 &	  &0.0741& 23.10&      &      &12   &   &   &		&\cr
SE	 &	  &	 & 22.90&      &      &$>$5 &   &   &		&\cr
\hline
1739+183 &UGC10944&   G	 &	&      &      &     & 2 &   &\cite{cf91362}&\cr
E	 &PGC60713&	 &	&      &      &     &   &   &		&\cr
W	 &	  &	 &	&      &      &     &   &   &		&\cr
\hline
1741+279 &4C27.38 &  Q	 &	& 26.75& 26.11&	    & 1 &   &\cite{mj92353,m9833210,m97125573}&\cr
NW	 &B2	  & 0.372&26.27	&      &      &16   &   &   &\cite{211}	&\cr
S(C-J)	 &	  &	 &25.78 &      &      &15   &   &   &		&\cr
\hline
1743+666 &8C	  &   G	 &	&25.58 &23.55 &     & 1 &   &\cite{lr92404}&\cr
NE	 &	  &0.272 &	&      &      &56   &   &   &		&\cr
SW(C-J)  &	  &	 &	&      &      &25   &   &   &		&\cr
\hline
1745+624 &	  &  Q	 &	&28.37 &28.11 &     & 1 &   &\cite{tv95345}&\cr
SW	 &	  &3.886 &	&      &      &0.053&   &   &		&\cr
\hline
1747+30  &B2	  &   G	 &	&23.96 &22.91 &	    & 1 &   &\cite{175,176}&\cr
N	 &	  &0.1297& 23.36&      &      &29   &   &   &		&\cr
\hline
1749+701 &	  &  BL  &	&27.57 &26.52 &	    & 1 &   &\cite{k981151295}&\cr
NW	 &	  & 0.77 & 	&      &      &0.030&   &   &		&\cr
\hline
1749+096 &OT081	  &  BL  &	&26.16 &25.83 &	    & 1 & C &\cite{l00364391}&\cr
N	 &	  & 0.322&26.11	&      &      &0.010&   &   &		&\cr
\hline
1751+681B&8C	  &   G  &	&      &      &     & 1 &   &\cite{lr92404}&\cr
SE	 &	  &	 &	&      &      &     &   &   &		&\cr
NW(C-J)	 &	  &	 &	&      &      &     &   &   &		&\cr
\hline
1752+32B &B2 	  &   G	 &	& 23.47& 22.30&	    & 2 &   &\cite{29,175,176}&\cr
SW	 &	  &0.0449& 23.13&      &      &31   &   &   &		&\cr
NE	 &	  &      & 22.87&      &      &21   &   &   &		&\cr
\hline
1753+580 &	  &   G	 &	&24.62 &$<$23.32&   & 1 & C &\cite{162} &\cr
SE	 &	  &0.160 &	&      &      &15   &   &   &		&\cr
SW(C-J)	 &	  &	 &	&      &      &15   &   &   &		&\cr
\hline
1754+631 &8C	  &  G	 &	&      &      &     & 1 &   &\cite{lr92404}&\cr
W	 &	  &	 &	&      &      &     &   &   &		&\cr
\hline
1754+292 &	  &  	 &	&      &      &	    & 1 &   &\cite{w98300790}&\cr
NE	 &        &      &	&      &      &     &   &   &		&\cr
\hline
1759+211 &PKS	  &   G	 &	&24.78 &23.11 &	    & 1 &   &\cite{29,214}&\cr
SW	 &4C21.51 &(0.08)&	&      &      &16   &   &   &		&\cr
\hline
1800+440 &OU401	  &  Q	 &	&26.56 &26.04 &     & 1 & C &\cite{pw981,xr99297,pb92655,k981151295}&\cr
SW	 &	  &0.663 &	&      &      &13   &   &   &		&\cr
\hline
1802+110 &3C368	  &  Q	 &	&27.37 &23.63 &     & 1 &   &\cite{b98299357}&\cr
NE	 &	  &1.132 &	&      &      &9.0  &   &   &		&\cr
SW(C-J)	 &	  &	 &	&      &      &19   &   &   &		&\cr
\hline
1803+784 &	  &  BL	 &	&27.34 &26.97 &     & 1 &   &\cite{k981151295}&\cr
W	 &	  & 0.68 &	&      &      &0.026&   &   &		&\cr
\hline
1803+661 &8C	  &  G	 &	&      &      &     & 2 &   &\cite{lr92404}&\cr
SE	 &	  &	 &	&      &      &     &   &   &		&\cr
NW	 &	  &	 &	&      &      &     &   &   &		&\cr
\hline
1807+279 &4C27.41 &   Q	 &	&27.66 &27.29 &	    & 1 &   &\cite{29,w98300790}&\cr
NE	 &B2	  &  1.76& 	&      &      &13   &   &   &		&\cr
\hline
1807+698 &3C371   &   BL &	& 24.84& 24.60&	    & 1 &   &\cite{al94247,pw981,140,160,181}&\cr
SW	 &	  & 0.050&23.43	&      &      &14   &   &   &\cite{29,192,198,268,nh484l107}&\cr
	 &	  &      &	&      &      &     &   &   &\cite{k981151295,m9912281,s99526643,g00530245}&\cr
\hline
1809+568 &7C	  &  	 &	&      &      &     & 1 &   &\cite{p0053490}&\cr
SE	 &CJ	  &      &	&      &      &     &   &   &		&\cr
\hline
1815+680 &4C68.20 &  G	 &	&25.09 &24.71 &     & 1 &   &\cite{lr92404}&\cr
SW	 &8C	  &0.230 &	&      &      &50   &   &   &		&\cr
\hline
1815+246 &	  &  	 &	&      &      &	    & 1 & C &\cite{w98300790}&\cr
E	 &        &      &	&      &      &     &   &   &		&\cr
\hline
1816+475 &4C47.48 &   Q	 &	& 27.73&26.24 &	    & 1 &   &\cite{gc91171,lb8763,7}&\cr
NW	 &	  & 2.225&27.34	&      &      &14   &   &   &		&\cr
\hline
1819+396 &	  &  G	 &	&26.85 &25.40 &     & 1 &   &\cite{df9527}&\cr
S	 &	  & 0.4	 &26.69	&      &      &0.50 &   &   &		&\cr
\hline
1820+689 &	  &  G	 &	&24.84 &23.65 &     & 1 & C &\cite{ol10841}&\cr
NW	 &	  &0.0880&	&      &      &58   &   &   &		&\cr
SW(C-J)	 &	  &	 &	&      &      &14   &   &   &		&\cr
\hline
1823+568 &OU539	  &  BL	 &	& 27.39& 26.88&	    & 1 & L &\cite{pw981,160,fc11195,k981151295}&\cr
S	 &4C56.27 & 0.664&	&      &      &11   &   &   &		&\cr
\hline
1826+796 &	  &  G	 &	&25.16 &25.33 &     & 1 &   &\cite{tv95345}&\cr
SW	 &	  &0.224 &	&      &      &0.036&   &   &		&\cr
\hline
1827+32  &B2	  &   G	 &	& 24.07& 23.08&     & 2 &   &\cite{175} &\cr
E	 &	  &0.0659& 22.80&      &      &70   &   &   &		&\cr
W	 &	  &      & 22.90&      &      &33   &   &   &		&\cr
\hline
1828+487*&3C380	  &   Q	 &	& 27.94&27.30 &     & 1 & L &\cite{wa9186,sr90140,255,p98294327,o991171143}&\cr
NW	 &	  & 0.692&	&      &      &2.6  &   &   &\cite{dvod97191,t98506637,k00521045}&\cr
\hline
1829+290 &	  &   G	 &	& 27.47& 25.26&     & 2 & L &\cite{df9527}&\cr
E	 &	  &0.842 &27.14	&      &      &0.25 &   &   &		&\cr
W	 &	  &	 &27.11	&      &      &0.25 &   &   &		&\cr
\hline
1830+285 &4C28.45 &   Q  &	&26.83 &26.25 &     & 1 &   &\cite{107,w98300790}&\cr
NW	 &B2	  &0.594 &	&      &      &52   &   &   &		&\cr
\hline
1832+280 &	  &  	 &	&      &      &	    & 1 &   &\cite{w98300790}&\cr
S	 &        &      &	&      &      &     &   &   &		&\cr
\hline
1833+326 &3C382	  &   S1 &	&25.34 &23.63 &     & 1 &   &\cite{bb92186,gf435116}&\cr
NE	 &B2	  &0.0578&23.55	&      &      &65   &   &   &		&\cr
\hline
1834+620 &TEX     &   G	 &	& 26.45& 23.78&     & 2 &   &\cite{l99348699,s00315395}&\cr
SE	 &        &0.5194&      &      &      &428  &   &   &		&\cr
NW	 &	  & 	 &	&      &      &442  &   &   &		&\cr
\hline
1842+681 &S4	  &  Q	 & 	&26.54 &26.29 &     & 1 &   &\cite{fc11195}&\cr
SE	 &	  &0.475 & 	&      &      &0.089&   &   &		&\cr
\hline
1842+455 &3C388   &   G	 &	& 25.73& 23.76&     & 1 & C &\cite{rb421l23,29,44}&\cr
SW	 &        &0.0908&23.71 &      &      &22   &   &   &		&\cr
NE(C-J)	 &	  & 	 &23.32	&      &      &7.9  &   &   &		&\cr
\hline
1845+797*&3C390.3 &   S1 &	&25.59 & 24.10&     & 1 &   &\cite{aw96376,pa96p35,p9749177,k981151295}&\cr
NW    \% &	  &0.0569&23.55 &      &      &99   &   &   &\cite{lp951097}&\cr
SE(C-J)	 &	  &	 &22.74	&      &      &54   &   &   &		&\cr
\hline
1846+326 &	  &  	 &	&      &      &	    & 1 & C &\cite{w98300790}&\cr
E	 &        &      &	&      &      &     &   &   &		&\cr
\hline
1849+670 &4C66.20 &  Q   &	&26.93 &26.52 &     & 1 &   &\cite{tv95345,fc11195}&\cr
NW	 &S4	  &0.657 &	&      &      &0.043&   &   &		&\cr
\hline
1850+702 &	  &   G	 &	& 24.32& 23.96&	    & 2 & C &\cite{162}	&\cr
NE	 &	  & 0.079&22.66 &      &      &33   &   &   &		&\cr
SE	 &	  & 	 &22.66	&      &      &33   &   &   &		&\cr
\hline
1856+737 &S5	  &  Q	 &	& 26.20&26.00 &     & 1 &   &\cite{tv10737}&\cr
NE	 &	  &0.460 &	&      &      &88   &   &   &		&\cr
\hline
1857+566 &4C56.28 &   Q	 &	& 27.57& 25.99&	    & 1 & B &\cite{gc91171,lb8763,7,29,170}&\cr
SE	 &        & 1.595& 27.28&      &      &61   &   &   &\cite{170,174}&\cr
\hline
1857+630 &B	  &   Q	 &	&      &      &	    & 1 &   &\cite{a982991159}&\cr
NE	 &        & 	 & 	&      &      &     &   &   &		&\cr
\hline
1901+319 &3C395	  &   Q  &	& 27.20&26.84 &	    & 1 & L &\cite{sm90503,l99352443}&\cr
NW	 &	  &0.635 &	&      &      &2.6  &   &   &		&\cr
\hline
1919+479 &4C47.51 &   G	 &	& 25.22& 23.19&     & 1 & L &\cite{29,41,42,46,47}&\cr
E	 &	  & 0.103& 23.21&      &      &260  &   &   &\cite{204,205}&\cr
\hline
1922+478 &	  &  Q	 &	&      &      &	    & 1 &   &\cite{pb92655}&\cr
N	 &	  &	 &	&      &      &	    &   &   &		&\cr
\hline
1923+210 &	  &  	 &	&      &      &	    & 1 &   &\cite{w98300790}&\cr
W	 &        &      &	&      &      &     &   &   &		&\cr
\hline
1924+507 &4C50.47 &   Q  &      & 27.04& 26.76&     & 1 &   &\cite{tv10737,170,174}&\cr
NE	 &	  & 1.098&  	&      &      &34   &   &   &		&\cr
\hline
1928+738*&4C73.18 &   Q	 &	& 26.58&26.49 &     & 1 & C &\cite{24,hs92489,kw901057,mb93298,14}&\cr
SE	 &	  & 0.302& 24.8	&      &      &46   &   &   &\cite{34,65,109,195,209}&\cr
NE(C-J)	 &	  &	 & 24.0	&      &      &---  &   &   &\cite{231,264,k981151295,t0053395,l0054166}&\cr
\hline
1928+681 &B       &   Q	 &	&      &      &	    & 1 &   &\cite{a982991159}&\cr
SE	 &	  & 	 &	&      &      &     &   &   &		&\cr
\hline
1939+605 &3C401   &   G	 &	& 26.37& 24.09&	    & 1 & C &\cite{ha288859,29,43}&\cr
SW	 &	  & 0.201&25.24 &      &      &23   &   &   &		&\cr
\hline
1940+50  &3C402N  &   G  &	& 24.31& 22.08&	    & 2 &   &\cite{29}	&\cr
	 &	  &0.0247&	&      &      &6.2  &   &   &		&\cr
\hline
1945+241 &	  &  	 &	&      &      &     & 1 &   &\cite{tg95238}&\cr
N	 &	  &	 &	&      &      &     &   &   &		&\cr
\hline	
1946+708 &	  &  G	 &	& 25.03&23.26 &     & 2 & S &\cite{tv95345,tv485l9,s9833370,p99521103}&\cr
SW	 &	  &0.101 &	&      &      &0.053&   &   &		&\cr
NE	 &	  &	 &	&      &      &0.020&   &   &		&\cr
\hline
1949+023 &3C403	  &  S2  &	&25.41 &22.59 &     & 1 &   &\cite{bb92186}&\cr
NW	 &	  &0.059 &23.99	&      &      &26   &   &   &		&\cr
\hline
1954+513 &OV591	  &   Q	 &	&27.55 &27.53 &     & 1 &   &\cite{kw901057,pw981,fc11195,b991221}&\cr
NW	 &	  &1.2230&26.68	&      &      & 43  &   &   &		&\cr
\hline
1957+405\%&3C405  &  S1	 &	& 27.73& 24.12&	    & 1 &   &\cite{23,24,kw93375,cb911691,cb9464}&\cr
NW       &Cyg A	  & 0.057& 24.57&      &      &47   &   &   &\cite{ka96p11,29,135,191,k98329873}&\cr
SE(C-J)	 &	  &	 & 24.00&      &      &26   &   &   &\cite{198,217,nv324888,k981151295,w0054427}&\cr
         &	  &      &      &      &      &     &   &   &\cite{c991182581}&\cr
\hline
2005-044 &3C407	  &  Q	 &	&26.65 &25.81 &     & 1 &L  &\cite{pg86365,bh9491}&\cr
NE	 &	  &0.589 &	&      &      &49   &   &   &		&\cr
\hline
2007+777 &S5	  &   BL &	& 25.96&26.20 &     & 1 &   &\cite{kw921687,mb93298,1,p00353937}&\cr
SW	 &	  &0.342 &	&      &      & 62  &   &   &\cite{fc11195}&\cr
\hline
2015+657 &	  &  Q	 &	&28.17 &27.83 &     & 1 &   &\cite{tv95345}&\cr
NW	 &	  &2.845 &	&      &      &0.081&   &   &		&\cr
\hline
2017+743 &4C74.25 &  Q   &	&      &      &     & 1 &   &\cite{fc11195}&\cr
E	 &S5	  &	 &	&      &      &     &   &   &		&\cr
\hline
2021+317 &	  &  	 &	&      &      &	    & 1 & S &\cite{w98300790}&\cr
NE	 &        &      &	&      &      &     &   &   &		&\cr
\hline
2025-218 &	  &  G	 &	&27.65 &24.88 &     & 1 &   &\cite{cr1091}&\cr
SW	 &	  &2.630 &	&      &      &11   &   &   &		&\cr
\hline
2033+187 &	  &  Q	 &	&      &      &     & 1 & L &\cite{m9833210}&\cr
N	 &	  &      &	&      &      &     &   &   &		&\cr
\hline
2036-254 &	  &  G	 &	&27.42 &24.29 &     & 1 &   &\cite{cr1091}&\cr
NE	 &	  &2.00	 &	&      &      &17   &   &   &		&\cr
\hline
2037+511 &3C418   &   Q	 &	& 28.41& 28.18&	    & 1 &B  &\cite{p951555,29,160,161}&\cr
NW	 &	  & 1.686&	&      &      &9.3  &   &   &		&\cr
\hline
2040-26  &	  &   G	 &	&      &      &	    & 2 &   &\cite{30,68}&\cr
SE	 &	  &	 &	&      &      &     &   &   &		&\cr
NW	 &	  &	 &	&      &      &     &   &   &		&\cr
\hline
2043+749 &4C74.26 &   Q	 &	& 25.30&24.71 &	    & 1 &   &\cite{rw901p,pb9213p}&\cr
SE	 &	  &0.104 &	&      &      &191  &   &   &		&\cr
\hline
2045+068 &3C424	  &   G	 &	&25.67 &23.60 &     & 2 &   &\cite{bb92186}&\cr
NW	 &	  &0.1270&24.34	&      &      &15   &   &   &		&\cr
SE	 &	  &	 &24.44	&      &      &6.7  &   &   &		&\cr
\hline
2054+611 &	  &	 &	&      &      &     & 1 &   &\cite{tv95345}&\cr
NE	 &	  &	 &	&      &      &     &   &   &		&\cr
\hline
2105+236 &	  &   G	 &	&27.63 &25.94 &     & 1 &   &\cite{cr1091}&\cr
NE	 &	  &2.479 &	&      &      &74   &   &   &		&\cr
\hline
2112+202 &	  &  	 &	&      &      &	    & 1 &   &\cite{w98300790}&\cr
NE	 &        &      &	&      &      &     &   &   &		&\cr
\hline
2116+26  &NGC7052 &   G  &	& 22.72& 22.10&	    & 2 &   &\cite{29,154,175,176}&\cr
N	 &B2	  &0.0164& 21.66&      &      &10   &   &   &		&\cr
S	 &	  &      & 21.88&      &      &10   &   &   &		&\cr
\hline
2116+818 &S5	  &  S1	 &	& 24.51&24.11 &     & 1 &   &\cite{tv10737}&\cr
NW	 &	  &0.086 &	&      &      &16   &   &   &		&\cr
\hline
2120+168 &3C432   &   Q  &	& 27.63& 25.64&     & 1 &   &\cite{bh94766,24,dtb289753}&\cr
SE	 &	  &1.805 & 25.74&      &      &28   &   &   &		&\cr
C-J	 &	  &	&$<$24.82&     &      &20   &   &   &		&\cr 
\hline
2121+248 &3C433   &  S2  &    	& 26.15& 22.76&	    & 1 &   &\cite{pc911960,bb92186,21,29,277}&\cr
N	 & 	  &0.1016&24.24	&      &      &30   &   &   &		&\cr
\hline
2128+048 &	  &  G	 &	&27.85 &25.46 &     & 1 &   &\cite{sod325943}&\cr
NW	 &	  &0.99  &	&      &      &0.090&   &   &		&\cr
\hline
2129+492 &	  &	 &	&      &      &     & 1 &   &\cite{tg95238}&\cr
SW	 &	  &	 &	&      &      &     &   &   &		&\cr
\hline
2131-021 &PKS	  &  BL  &	&26.87 &26.82 &     & 1 &   &\cite{sd325911}&\cr
SE	 &	  &0.557 &	&      &      &14   &   &   &		&\cr
\hline
2134+004 &PKS	  &  Q	 &	&28.50 &27.37 &	    & 1 & S &\cite{l0054166}&\cr
NW	 &        & 1.932&	&      &      &0.014&   &   &		&\cr
\hline
2138+826 &S5	  &  Q	 &	&27.95 &27.51 &	    & 1 &   &\cite{w98300790}&\cr
SE	 &        & 2.35 &	&      &      &6.4  &   &   &		&\cr
\hline
2141+279 &3C436	  &  G	 &	&26.21 &23.99 &     & 1 &   &\cite{ha288859}&\cr
S	 &	  &0.2145&	&      &      &107  &   &   &		&\cr
\hline
2145+067 &4C06.69 &  Q	 &	&28.19 &27.84 &     & 1 &   &\cite{fc11195}&\cr
SE	 &PKS	  &0.99  &	&      &      &0.038&   &   &		&\cr
\hline
2147+145 &        &  Q	 &	&      &      &     & 1 & L &\cite{m9833210}&\cr
NE	 &	  &      &	&      &      &     &   &   &		&\cr
\hline
2148-555 &B	  &  G	 &	&26.14 &22.91 &     &   &   &\cite{l96p227}&\cr
NE	 &	  &0.035 &	&      &      &101  &   &   &		&\cr
SW	 &	  &	 &	&      &      &132  &   &   &		&\cr
\hline
2149+212 &4C21.59 &  Q	 &	&27.55 &26.56 &     & 1 &   &\cite{lb8763}&\cr
SW	 &	  &1.534 &	&      &      &7.6  &   &   &		&\cr
\hline
2149-158AB&PKS	  &   G	 &	&24.26 &22.80 &     & 2 & C &\cite{pc911960,ol10841}&\cr
E	 &	  &0.0616&	&      &      &71   &   &   &		&\cr
NW	 &	  &	 &	&      &      &51   &   &   &		&\cr
\hline
2150+173 &PKS	  &  BL  &	&      &      &     & 1 &   &\cite{fc11195}&\cr
W	 &	  &	 &	&      &      &     &   &   &		&\cr
\hline
2152+085 &	  &   G	 &	&24.77 &23.28 &     & T & C &\cite{ol10841}&\cr
NW	 &	  &0.1500&	&      &      &26   &   &   &		&\cr
SW(C-J)	 &	  &	 &	&      &      &29   &   &   &		&\cr
\hline
2152-699 &PKS	  &   G	 &	&25.82 &23.86 &     & 2 &   &\cite{f98296701}&\cr
NE	 &	  &0.0282&	&      &      &14   &   &   &		&\cr
SW	 &	  &	 &	&      &      &2.4  &   &   &		&\cr
\hline
2153+377 &3C438   &   G	 &	& 26.86& 23.99&	    & 2 &   &\cite{lp91537,ha288859,24,29}&\cr
NW	 &	  & 0.292& 25.19&      &      &24   &   &   &		&\cr
SE(C-J)	 &	  &      & 25.11&      &      &24   &   &   &		&\cr
\hline
2200+420*&BL Lac  &  BL  &	&25.77 &25.07 &     & 1 & L &\cite{pw981,fc11195,t98500810,d0012961}&\cr
SW	 &	  &0.0688&	&      &      &0.12 &   &   &		&\cr
\hline
2201+315 &4C31.63 &   Q	 &	&26.25 &26.18 &     & 1 &   &\cite{mb93298,fc11195,b991221,l0054166}&\cr
SW	 &GC	  &0.298 &	&      &      &102  &   &   &		&\cr
	 &B2	  &	 &	&      &      &     &   &   &		&\cr
\hline
2201+044 &4C04.77 &  BL  &	& 24.10&23.41 &     & 1 &   &\cite{lmk93875,s99526643}&\cr
NW	 &PKS	  &0.028 &	&      &      & 2   &   &   &		&\cr
\hline
2202+128 &	  &   G	 &	&27.37 &24.82 &     & 1 &   &\cite{cr1091}&\cr
SW	 &	  &2.704 &	&      &      &8.4  &   &   &		&\cr
\hline
2203+292 &3C441	  &   G	 &	&26.96 &$<$23.16&   & 1 &   &\cite{l98298966}&\cr
SE	 &	  & 0.708&	&      &      &39   &   &   &		&\cr
\hline
2206+650 &	  &  	 &	&      &      &     & ? &   &\cite{pb92655}&\cr
W	 &	  &	 &	&      &      &     &   &   &		&\cr
\hline
2209+152 &MC3     &   Q	 &	& 26.85&25.98 &	    & 1 &   &\cite{gc91171,lb8763,7}&\cr
NE	 &	  & 1.502&26.63	&      &      &23   &   &   &		&\cr
\hline
2209+080 &4C08.64 &   Q	 &	&26.52 &25.83 &     & 1 &   &\cite{pg86365}&\cr
S	 &PKS	  &0.484 &	&      &      &25   &   &   &		&\cr
\hline
2210+016 &	  &  G?  &	&26.92 &25.72 &     & 1 &   &\cite{sod325943}&\cr
W	 &	  &(0.5) &	&      &      &0.20 &   &   &		&\cr
\hline
2212+136 &NGC7237 &   G	 &	&24.41 &22.10 &     & ? &   &\cite{cf91362}&\cr
NE	 &UGC11958&0.0268&	&      &      &63   &   &   &		&\cr
\hline
2213-283 &MRC	  &   Q	 &	&27.00 &25.83 &     & 1 &   &\cite{k98118327}&\cr
E	 &B2	  &0.946 &	&      &      &210  &   &   &		&\cr
\hline
2214+358A&MGC     &   Q	 &	&26.24 &24.90 &     & 1 &   &\cite{m984929}&\cr
E	 &	  &0.879 &	&      &      &12   &   &   &		&\cr
\hline
2221-023 &3C445	  &  S1	 &	& 25.30& 23.51&	    & 1 &   &\cite{29,lb29120}&\cr
S	 &	  & 0.057&23.04 &      &      &210  &   &   &		&\cr
N(C-J)	 &	  &	 &22.15 &      &      &33   &   &   &		&\cr
\hline
2223-052 &3C446	  &  BL  &	& 28.15&28.20 &     & 1 &   &\cite{fp92459,k981151295}&\cr
E	 &	  &1.404 &	&      &      &1.4  &   &   &		&\cr
\hline
2229-086 &A2448AB &   G  &	& 24.68&23.23 &     & 2 &   &\cite{ow80501,ol10841}&\cr
E	 &	  &0.073 &	&      &      &96   &   &   &		&\cr
W	 &	  &	 &	&      &      &102  &   &   &		&\cr
\hline
2229+391 &3C449   &   G  &	& 24.02& 22.07&	    & 2 & B &\cite{cm9461,cf91362,20,29,49}&\cr
N	 &UGC12064&0.0171&$<$22.1&     &      &19   &   &   &\cite{67,112,h982961098,f9934129} &\cr
S	 &B2	  &      &  	&      &      &19   &   &   &		&\cr
\hline
2230+114 &CTA102  &  Q	 &	& 28.04&27.68 &     & 1 & L &\cite{w89p529,fc11195,k981151295,t0053395}&\cr
SE	 &4C11.69 &1.037 &	&      &      &0.11 &   &   &		&\cr
\hline
2231+359 &B       &  	 &	&      &      &     & 1 &   &\cite{a982991159}&\cr
NE	 &	  &	 &	&      &      &     &   &   &		&\cr
\hline
2236+35  &B2      &   G	 &	& 23.44& 21.83&	    & 2 & S &\cite{29,154,175,176}&\cr
NE	 &        &0.0277& 22.83&      &      &9    &   &   &		&\cr
SW	 &	  &      & 22.88&      &      &9    &   &   &		&\cr
\hline
2236-176 &PKS	  &   G	 &	& 24.96&22.71 &     & 2 & L &\cite{68,167}&\cr
SW	 &	  &0.0698&24.12	&      &      &37   &   &   &		&\cr
NE	 &	  &	 &24.12	&      &      &37   &   &   &		&\cr
\hline
2240-260 &PKS	  &  BL  &	&26.87 &26.73 &     & 1 & C &\cite{sd325911}&\cr
NE	 &	  &0.774 &	&      &      &30   &   &   &		&\cr
\hline
2243+394 &3C452	  &   G	 &	&25.90 &24.23 &     & 1 &   &\cite{bb92186}&\cr
SW	 &	  &0.0811&23.84	&      &      &74   &   &   &		&\cr
NE(C-J)	 &	  &	 &23.57	&      &      &74   &   &   &		&\cr
\hline
2243-178 &A2480	  &   G	 &	&24.89 &23.28 &     & 2 &   &\cite{bb9353}&\cr
N	 &	  &0.1234&	&      &      &19   &   &   &		&\cr
S	 &	  &	 &	&      &      &18   &   &   &		&\cr
\hline
2248+192 &4C19.74 &  Q	 &	&27.55 &25.82 &     & 1 & S &\cite{lb8763}&\cr
NW	 &	  &1.806 &	&      &      &13   &   &   &		&\cr
\hline
2249+185 &3C454	  &   Q	 &	&28.04 &$<$27.02&   & 1 & L &\cite{ss91225,l98299467}&\cr
S	 &	  &1.76	 &27.74 &      &      &	2.3 &   &   &		&\cr
N(C-J)	 &	  &	 &26.57	&      &      &0.46 &   &   &		&\cr
\hline
2251+134 &4C13.85 &   Q  &      & 26.88& 26.51&	    & 1 &   &\cite{pg86365,211}&\cr
SE	 &	  & 0.673&	&      &      &15   &   &   &		&\cr
\hline
2251+158*&3C454.3 &   Q	 &	& 28.10& 28.03&	    & 1 &   &\cite{c9051,cg96861,35,179,198}&\cr
NW	 &	  & 0.859& 26.43&      &      &21   &   &   &\cite{29,36,192,195,ka96p11}&\cr
	 &	  &      &      &      &      &     &   &   &\cite{k981151295,g9952274,l00364391}&\cr
\hline
2252-089 &PKS	  &	 &	&      &      &     & 1 &B,S&\cite{fc11195}&\cr
SW	 &	  &	 &	&      &      &     &   &   &		&\cr
\hline
2252+129 &3C455	  &   Q	 &	&26.79 &23.76 &     & 1 &   &\cite{bh9491}&\cr
SW	 &	  &0.555 &	&      &      &4.3  &   &   &		&\cr
\hline
2259+137 &	  &  G	 &	& 24.91&23.74 &     & 2 & C &\cite{ol10841}&\cr
NE	 &	  &0.1721&	&      &      &50   &   &   &		&\cr
NW	 &	  &	 &	&      &      &78   &   &   &		&\cr
\hline
2259+568 &	  &      &	&      &      &	    & 1 &   &\cite{pb92655}&\cr
SE	 &	  &      &	&      &      &     &   &   &		&\cr
\hline
2300-189 &PKS     &   Q  &  	& 25.44& 24.90&     & 1 & S &\cite{29,111}&\cr
NW	 & 	  & 0.129&  	&      &      &68   &   &   &		&\cr
NE(C-J)	 &        & 	 &	&      &      &68   &   &   &		&\cr
\hline
2305+18  &PKS	  &   Q	 &	& 26.08& 24.78&	    & 1 &   &\cite{93,94}&\cr
N	 &4C18.68 & 0.313&	&      &      &4.0  &   &   &		&\cr
\hline
2308+072 &4C07.61 &  G	 &	&24.75 &24.10 &     & 2 & C &\cite{vb43667}&\cr
NE	 &NGC7503 &0.0448&23.19	&      &      &30   &   &   &		&\cr
SW(C-J)	 &	  &	 &22.89	&      &      &30   &   &   &		&\cr
\hline
2309+184 &3C457	  &   G	 &	& 26.64&24.21 &     & 1 &   &\cite{lp91537}&\cr
NE	 &PKS	  &0.428 &	&      &      & 10  &   &   &		&\cr
\hline
2314-116 &PKS	  &   Q	 &	&25.99 &25.56 &     & 1 &   &\cite{pg86365}&\cr
N	 &	  &0.549 &	&      &      &48   &   &   &		&\cr
\hline
2316+184 &OZ127   &   G	 &	& 23.70& 22.41&	    & 2 & C &\cite{29,162}&\cr
NE	 &	  &0.0395&	&      &      &21   &   &   &		&\cr
SE	 &	  &	 &	&      &      &21   &   &   &		&\cr
\hline
2318+079 &NGC7626 &   G  &	& 23.17& 21.31&     & 2 &   &\cite{cf91362,19,29}&\cr
SW	 &UGC12531&0.0112&	&      &      &7.5  &   &   &		&\cr
SE(C)	 &	  &	 &	&      &      &7.5  &   &   &		&\cr
\hline
2318+620 &	  &	 &	&      &      &     & 2 &   &\cite{tg95238}&\cr
N	 &	  &	 &	&      &      &     &   &   &		&\cr
S	 &	  &	 &	&      &      &     &   &   &		&\cr
\hline
2320+506 &	  &	 &	&      &      &     & 1 & L &\cite{fc11195}&\cr
SW	 &	  &	 &	&      &      &     &   &   &		&\cr
\hline
2322+143A&	  &  G	 &	&24.22 &22.25 &     & 2 & C &\cite{162}&\cr
NE	 &	  &0.0440&23.26	&      &      &3.6  &   &   &		&\cr
SW	 &	  &	 &23.26	&      &      &3.6  &   &   &		&\cr
\hline
2325+29  &4C29.68 &   Q  &	& 27.34& 26.37&     & 1 &   &\cite{gc91171,29,253}&\cr
SE	 &	  & 1.015&27.03	&      &      &85   &   &   &		&\cr
\hline
2330+091 &	  &  G	 &	&24.82 &22.56 &     & 2 & C &\cite{ol10841}&\cr
N	 &	  &0.1623&	&      &      &17   &   &   &		&\cr
SE	 &	  &	 &	&      &      &26   &   &   &		&\cr
\hline
2335+267 &3C465   &   G  &	& 24.85& 23.37&     & 1 & L &\cite{cf91362,vc454735,29,41,66}&\cr
NW	 &NGC7720 &0.0293& 	&      &      &24   &   &   &\cite{67,131,166,d0012933}&\cr
SE(C-J)	 &UGC12716&	 &	&      &      &     &   &   &		&\cr
\hline
2335-027 &PKS	  &  Q   &	&27.39 &26.63 &     & 1 & L &\cite{fc11195}&\cr
NE	 &	  &1.072 &	&      &      &0.19 &   &   &		&\cr
\hline
2337+268 &NGC7728 &   G  &	& 23.49& 23.15&     & 2 &   &\cite{cf91362,29}&\cr
NE	 &UGC12727&0.0314&      &      &      &39   &   &   &		&\cr
SW	 &	  &	 &	&      &      &39   &   &   &		&\cr
\hline
2338+042 &4C04.81 &   Q  &	& 27.98& 27.15&     & 1 &   &\cite{lb8763,7,29}&\cr
SE	 &	  & 2.594& 27.24&      &      &4.6  &   &   &		&\cr
\hline
2338-290 &	  &   Q  &	& 26.04& 25.07&     & 1 & C &\cite{i98300269}&\cr
NE	 &	  & 0.446& 	&      &      &86   &   &   &		&\cr
\hline
2346+753 &	  &  	 &	&      &      &	    & 1 & S  &\cite{w98300790}&\cr
SE	 &        &      &	&      &      &     &   &   &		&\cr
\hline
2349+327 &4C32.69 &   Q  &	& 27.53& 25.15&	    & 1 &   &\cite{pg86365,29,78,bh9491}&\cr
NW	 &        & 0.671& 25.69&      &      &127  &   &   &		&\cr
SE(C-J)  &	  &	 &	&      &      &76   &   &   &		&\cr
\hline
2352+495 &OZ488	  &  G	 &	&26.26 &25.74 &     & 1 &   &\cite{pw981,tr46395}&\cr
S(C-J)	 &	  &0.2383&24.45	&      &      &0.097&   &   &\cite{rt460612}&\cr
N	 &	  &	 &24.28 &      &      &0.057&   &   &		&\cr
\hline
2353+283 &4C28.59 &  Q	 &	& 26.74&25.58 &     & 1 &   &\cite{pg86365}&\cr
SW	 &OZ289	  &0.731 &	&      &      &25   &   &   &		&\cr
\hline
2354+471 &4C47.63 &   G  &	& 24.63& 22.49&	    & 1 &   &\cite{29,46,248}&\cr
NE	 &	  &0.046&$>$21.56&     &      &37   &   &   &		&\cr
SW(C)	 &	  &	 &	&      &      &     &   &   &		&\cr
\hline
2354+144 &4C14.85 &  Q	 &	& 27.72&26.57 &     & 1 &   &\cite{lb8763}&\cr
SE	 &	  &1.810 &	&      &      &24   &   &   &		&\cr
\hline
2356+701 &	  &	 &	&      &      &	    & 1 &   &\cite{pb92655}&\cr
N	 &	  &	 &	&      &      &	    &   &   &		&\cr
\hline
2357+004 &PKS	  &   G	 &	&25.32 &23.65 &     & 1 &   &\cite{pc911960}&\cr
N	 &	  &0.0839&	&      &      &36   &   &   &		&\cr
S(C-J)	 &	  &	 &	&      &      &24   &   &   &		&\cr
\hline
2359-159 &PKS	  &   G	 &	&      &      &     & 2 &   &\cite{pc911960}&\cr
NW	 &	  &	 &	&      &      &     &   &   &		&\cr
SE	 &	  &	 &	&      &      &     &   &   &		&\cr
\hline
}
\vskip 1pc
*: Superluminal sources; \%: Subluminal sources

\end{document}